\documentclass[twocolumn,preprintnumbers,amsmath,amssymb,aps,pre,reprint,longbibliography]{revtex4-1}
\usepackage{graphicx}
\begin{document}

\title{
Collective Effects and Pattern Formation For Directional Locking of Disks Moving Through Obstacle Arrays 
} 
\author{
C. Reichhardt and C. J. O. Reichhardt 
} 
\affiliation{
Theoretical Division and Center for Nonlinear Studies,
Los Alamos National Laboratory, Los Alamos, New Mexico 87545, USA\\ 
} 

\date{\today}
\begin{abstract}
We examine directional locking effects in
an assembly of disks driven through a square array
of obstacles as the angle of drive rotates from zero to ninety degrees.
For increasing disk densities, the system exhibits a series of
different dynamic patterns along certain locking directions,  
including one-dimensional or multiple row chain phases
and density modulated phases.  
For non-locking driving directions, the disks form disordered patterns
or clusters.
When the obstacles are small or far apart,
a large number of locking phases appear; however,
as the number of disks increases, the number of
possible locking phases drops due to
the increasing frequency of collisions between the disks and obstacles.  
For dense arrays or large obstacles,
we find an increased clogging effect in which immobile and moving
disks coexist.
\end{abstract}
\maketitle

\section{Introduction}

In directional locking,
the motion of a particle
driven over a periodic substrate such as a square 
lattice tends to remain locked at certain symmetry directions
of the lattice
as the driving direction is varied
\cite{Reichhardt99,Wiersig01,Korda02,Gopinathan04,MacDonald03,Balvin09}.
For a square lattice,
the strongest directional locking appears when the
angle $\theta$ between the driving direction and the lattice
symmetry direction is
near $\theta=0^\circ$, $45^\circ$, or $90^\circ$; however, locking can occur
for any rational ratio $p/q$ with integer $p$ and $q$,
where the particle moves exactly $p$ lattice constants
in the $x$ direction and $q$ lattice constants in the $y$
direction during some time interval.
In this case, the directional locking is centered at an angle
$\theta_L = \arctan(p/q)$,
so that $p/q = 0$ corresponds to 
$\theta_L=0^\circ$ and $p/q = 1$
gives $\theta_L=45^{\circ}$.
Locking occurs whenever $\theta=\theta_L \pm \Delta\theta$,
where the width of an individual locking step is defined to be $2\Delta\theta$,
and the step width is
largest 
for small values of $p$ and $q$.
This behavior is similar to the Bragg
angle scattering conditions found for a lattice.
A devil's staircase hierarchy of
locking steps can appear in 
the angle of the particle motion as a function of
the driving angle $\theta$.

Directional locking effects have similarities to
the phase locking 
phenomena observed
in systems with two competing frequencies.
For example, if a particle moving over a one-dimensional (1D)
periodic substrate under
a dc drive is subjected to an additional ac drive in the direction of
motion,
locking effects arise
due to the interaction between
the frequency $\omega_1$ generated by the dc motion of the particle 
over the substrate and
the ac driving frequency $\omega_2$.
When $\omega_2$ is fixed,
the particle velocity locks to a constant value over a range of
dc drive amplitudes in order to maintain resonance
\cite{Shapiro63,Coppersmith86,Barone82}.
Varying the dc drive changes
$\omega_{1}$,
so
a series of steps known as Shapiro steps appear
in the velocity-force curve
at rational values of
$\omega_1/\omega_2$
\cite{Shapiro63,Coppersmith86,Barone82}.
Directional locking was first studied 
for vortices in type-II superconductors
driven in a changing direction
over a periodic pinning lattice,
and occurs when the direction of vortex motion becomes locked to
one of the substrate symmetry
directions
\cite{Reichhardt99}.
The substrate lattice structure determines the set of
possible directional locking angles, while   
the widths of the locking steps can be changed by varying
the strength of the pinning.
The lattice over which the superconducting vortices move
is composed of
attractive pinning sites
which the vortices must traverse
during a locking step.
Due to the
relatively long range of the vortex-vortex interactions,
the vortices form a moving lattice or ordered structure
on the locking steps,
whereas
when the motion is not locked,
the vortex structure is more disordered
or liquid-like \cite{Reichhardt99,Reichhardt12}.

Following the vortex work,
similar directional locking effects were proposed
to occur for classical electrons
moving over a square antidot lattice \cite{Wiersig01}.
Here, the direction of drive is controlled by the applied magnetic field, since
a larger magnetic field produces a larger Hall angle.
In the
presence of a periodic array of scattering sites,
the Hall angle becomes quantized with steps at rational values of
$p/q$, where the electron translates by an integer number $p$ and $q$ of
substrate lattice constants in the $x$ and $y$ directions, respectively,
during a period of time.

The first experimental observations of directional lockings
were obtained using
colloidal particles moving 
over a square optical trap array.
When the driving direction is fixed but the substrate lattice is
rotated,
the colloidal motion locks to different symmetry angles \cite{Korda02}.
The width $2\Delta\theta$ of the locking steps
depends on the interaction of the particle with the
pinning site or obstacle,
implying that if two species of particles are present that each
interact differently with the substrate,
the driving conditions can be tuned such that one species locks to the
substrate and the other does not.
As a result, each species moves at a different angle,
making it possible to achieve a spatial
separation or fractionation of the species.
This sorting effect based on directional locking
was first demonstrated experimentally
by MacDonald {\it et al.}
for colloids of
different sizes and
different refractive indexes \cite{MacDonald03}.
Directional locking and sorting effects 
have been studied extensively
for colloidal systems under different conditions 
and geometries, taking into
account a variety of dynamical effects and particle-substrate interactions 
\cite{MacDonald03,Huang04,Reichhardt04b,Pelton04,Ladavac04,Lacasta05,Gleeson06,Roichman07a,Long08,Koplik10,Speer10,Risbud14,Wunsch16,Tran17,Chen18,Li18,Li19}. 
Directional locking for particles
on periodic substrates has also been studied
in the context of skyrmions in chiral magnets.
Skyrmions are particle-like magnetic textures or bubbles
which
exhibit a Hall angle that changes with increasing driving force
in the presence of a substrate \cite{Reichhardt15a,Feilhauer19,Vizarim20}.      
There are also a number of systems
in which particles moving over periodic substrates under
different driving conditions
preferentially move in certain directions due to the underlying
symmetry of the substrate.
This effect has been studied for magnetic colloids undergoing oscillatory 
motion 
\cite{Soba08,Loehr16}
and for active matter on periodic obstacle arrays
\cite{Volpe11,BrunCosmeBruny20}. 

In many studies of directional locking, the dynamics is effectively
in the single particle limit;
however, when collective effects become important,
such as when
a large number of interacting particles are present,
changes in the locking behavior can occur
that include modifications of
locking step widths or the formation of different types of patterns
in locking and non-locking regimes.
Collective effects of this type were
studied for superconducting vortices
on a periodic pinning array, and they can also appear
for vortices or colloids moving over quasiperiodic arrays,
where for certain driving
directions the system forms moving smectic, square, triangular lattice, 
or disordered phases \cite{Reichhardt11}. 
Several of these phases have been observed in experiments 
on colloids moving over quasiperiodic pinning arrays \cite{Bohlein12a}. 
In the superconducting vortex system,
the directional locking step widths oscillate as a function of magnetic
field, with wider steps or stronger locking occurring
when the number of vortices is an integer multiple
of the number of
pinning sites \cite{Silhanek03,Villegas03,Reichhardt08b,Zechner18}. 

More recently, 
the impact of collective effects on locking phases was studied
for colloidal clusters where each individual cluster can have different
orientations that lock to the orientation of the underlying
substrate lattice \cite{Tierno19,Cao19}.
Similar effects were observed
for the motion of Au islands on two-dimensional (2D) atomic
substrates \cite{Trillitzsch18}.  
Recent studies focused on collective locking effects for colloids moving 
over triangular substrate arrays, where certain driving directions have
two equivalent locking directions.
If the particles do not interact with each other,
no net directional locking occurs,
but when particle-particle interactions
are added,
a global locking effect appears due to a dynamical symmetry breaking.
The direction of this symmetry breaking
can be controlled
using a small biasing field \cite{Stoop20}.
Similar spontaneous dynamic symmetry breaking leading to directional
locking has also been found
in simulations of
colloids \cite{Reichhardt04} and vortices \cite{Reichhardt10}
on periodic substrate
arrays.   

Up until now,
collective locking on pinning arrays has been studied for
particles with relatively long range interactions, such as superconducting
vortices or charged colloids.
Far less is known about the impact of collective effects
on locking when
the particles have short range interactions and are
moving over arrays of repulsive obstacles or posts.
A limited study addressed
the dense particle limit of bidisperse disks moving
through
a square obstacle array,
and showed that a
clogged state can appear in which the disk density becomes
strongly inhomogeneous
and the disks pile 
up behind each other \cite{Nguyen17}.
In this case, the clogging
susceptibility depends on the drive angle, with clogging occurring much
more readily at certain driving angles.
    
In this work we study directional locking for
a monodisperse assembly of disks moving through a square array of
obstacles 
where we vary the density of the moving disks as well as the
radius and lattice constant of the obstacles.
For low disk densities and small obstacle sizes,
strong directional locking occurs and
the system is nearly always in a locked state with $\theta_L = \arctan(p/q)$.
As the disk density increases,
fewer locking steps appear
and the disks form 1D chains at
certain driving directions.  For higher disk densities,
the chains thicken and we find that two to three rows of disks
can move between adjacent obstacles.
When an integer number of rows of moving disks is unable to form,
the locking effects
are lost.
This is similar to the frustration effect
found for the ordering of rows of disks on quasi-1D substrates. 
On other locking steps,
the system forms a density modulated state
where the overall configuration
is disordered but the trajectories of individual disks
are ordered.
In the non-locking regimes, the 
disk configurations are more disordered
and
mixing of the disks occurs.
The velocity of the disks in a locking region
is not constant 
but changes non-monotonically with driving angle $\theta$, exhibiting
peaks and valleys.
In contrast,
the direction of disk motion is constant on a locking step.
There are pronounced cusps and dips in the disk velocity
at the transitions into and out of the locking phases.
At high disk densities, we find that the
$p/q = 1.0$ locking step is lost due to a
dynamical frustration effect,
but some other locking phases remain present. 
When the obstacle size is increased in a system with fixed
disk density,
the width of the locking phases varies non-monotonically
and a clogging effect emerges in
which a portion of the disks
become blocked behind the obstacles.
The amount of clogging that occurs depends on the
direction of driving, and the disks are able to slide more easily when
the drive is aligned with
$0^{\circ}$, $90^{\circ}$, or $45^{\circ}$.
For the largest obstacles or for high disk densities,
we find a complete clogging where all disk flow ceases.  
When the disk density and obstacle size are held constant but the obstacle
lattice constant is increased,
the number of locking steps
increases but the width of the locking phases is reduced. 
Our results should be relevant to the flow of uncharged colloids,
bubbles or emulsions over obstacle arrays,
and suggest new
ways to generate different dynamical patterns.  

\section{Simulation} 
We consider a two-dimensional system
of size $L \times L$
with periodic boundary conditions in the $x$- and $y$-directions
where $L=36$.
The sample contains
a square lattice of
$N_{\rm obs}$ obstacles 
modeled as harmonically repulsive posts
with radius 
$R_{\rm obs}$ and lattice constant $a$, as well as 
$N_{d}$ harmonically repulsive disks of radius $R_d$.
The overall system density $\phi$
is defined to be the total area covered by both obstacles
and disks, $\phi =  (N_{\rm obs}\pi R^{2}_{\rm obs} + N_{d}\pi R^{2}_{d})/L^2$. 
The dynamics of disk $i$ is obtained
by integrating the following overdamped equation of motion: 
\begin{equation} 
\alpha_d {\bf v}_{i}  =
{\bf F}^{dd}_{i} + {\bf F}^{\rm obs}_{i} + {\bf F}^{D} .
\end{equation}
Here
${\bf r}_{i}$ is the disk position,  
${\bf v}_{i} = {d {\bf r}_{i}}/{dt}$ is
the disk velocity,
and $\alpha_d$ is the damping constant which we set to
$\alpha_d=1.0$. 
The disk-disk interaction force is
${\bf F}^{dd}_{i}$ and the disk-obstacle
interaction force is ${\bf F}^{\rm obs}$, while  
the driving force is
${\bf F}^{D} = F^{D}\cos(\theta){\bf \hat{x}} + F^{D}\sin(\theta){\bf \hat{y}}$.
We gradually increase $\theta$ from zero so that
${\bf F}^{D}$
is initially aligned with the
$x$-direction and rotates into the $y$ direction.
We measure the average velocity in the $x$- and $y$- directions, 
$\langle V_{x}\rangle = \sum^{N^{d}}_{i= 1}{\bf v}_i\cdot {\bf {\hat x}}$   
and $\langle V_{y}\rangle = \sum^{N_{d}}_{i= 1}{\bf v}_i\cdot {\bf {\hat y}}$,
as well as the net velocity
$\langle V\rangle = \sqrt{\langle V_x\rangle^{2} + \langle V_y\rangle^{2}}$.
The drive is fixed to $F^D=0.5$ and we increment $\theta$ by
an amount
$\Delta \theta = 0.057^\circ$ every
$10^4$ simulation time steps.
We have also used slower increment rates
for the smaller obstacle radii
in order to resolve the higher order directional locking effects.   

\section{Varied Disk Density}

\begin{figure}
\includegraphics[width=\columnwidth]{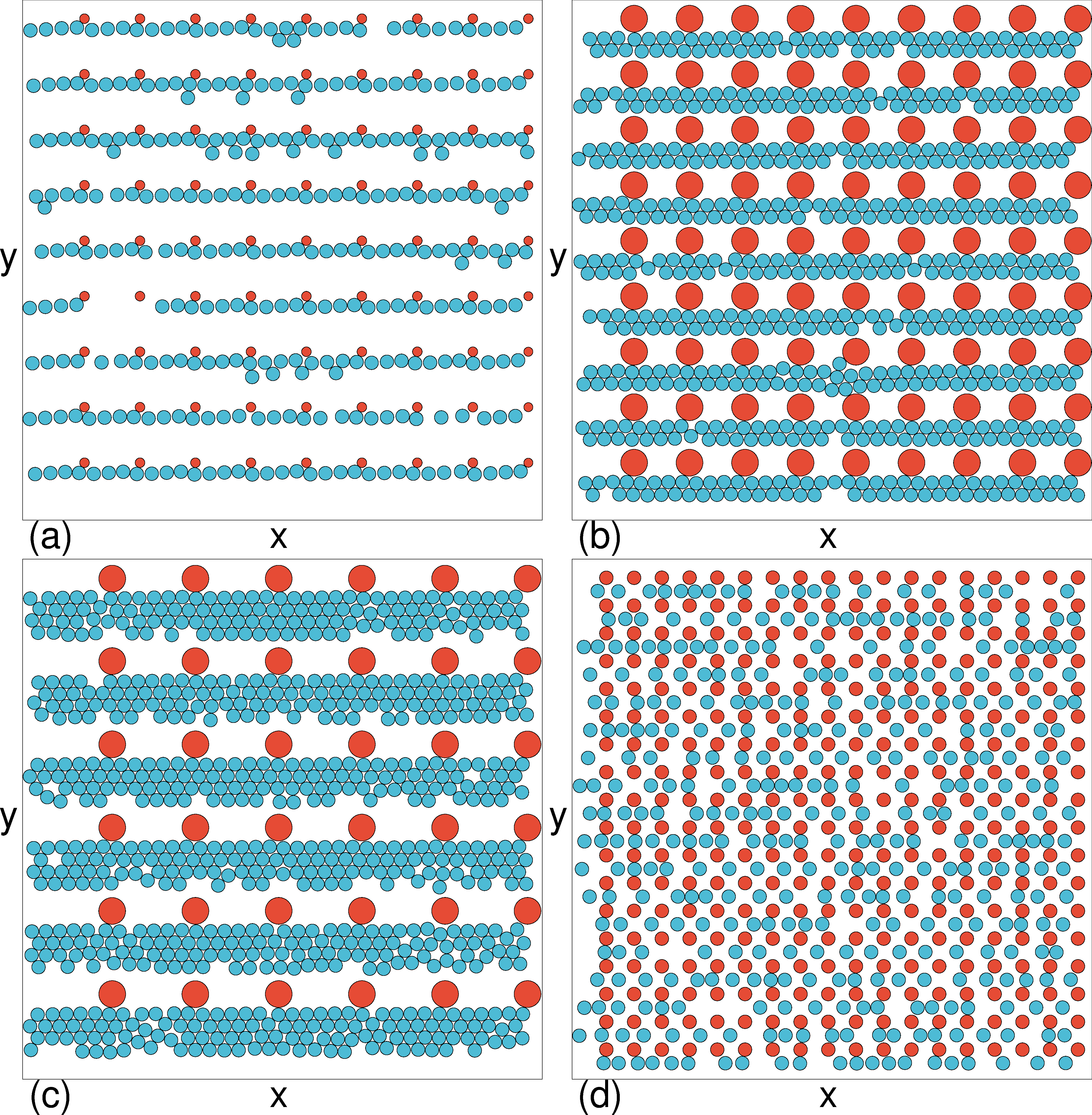}
\caption{
The obstacle locations (red) and the mobile disks (blue)
for a system with disk radius 
$R_d=0.5$
and driving force $F^D=0.5$
applied at
an angle $\theta = 3^\circ$ from the positive $x$ axis.
The disk motion is locked to $\theta_L = 0^\circ$. 
(a) Obstacle lattice constant $a = 4.0$ and radius $R_{\rm obs} = 0.35$ at
a total disk density of $\phi = 0.217$.
(b) For $a=4.0$, 
$R_{\rm obs} = 1.0$,
and a
larger number of mobile disks giving
$\phi = 0.57$, there are two mobile rows of disks between each
row of obstacles.
(c) $a = 6.0$,
$R_{\rm obs} =1.0$,
and $\phi = 0.55$,
where a density modulated state appears.
(d) $a=2.0$,
$R_{\rm obs} = 0.5$,
and $\phi = 0.242$. 
}
\label{fig:1}
\end{figure}

In Fig.~\ref{fig:1} we illustrate the positions of
the disks and obstacles
for a disk radius of $R_d=0.5$ when the driving angle has
reached $\theta=3^\circ$.
In the absence
of a substrate the disks would move along the $\theta$ direction,
but in the presence 
of obstacles the motion locks to $\theta_L=0^\circ$, corresponding to
$p/q=0$,
due to the symmetry of the 
substrate lattice.
In Fig.~\ref{fig:1}(a), the obstacle lattice constant is
$a = 4.0$, the obstacle radius is $R_{\rm obs} = 0.35$, and
the sample contains $N_{\rm obs} = 81$ obstacles
and $N_{d} = 319$ disks for a total density of  
$\phi = 0.217$.
Here the disks form nearly 1D chains
of single rows which 
brush up against the obstacles due to the nonzero driving angle.
In Fig.~\ref{fig:1}(b),
we show the same system
for a larger obstacle
radius
$R_{\rm obs} = 1.0$
and larger number of disks
$N_{d} = 619$, giving a total density of
$\phi = 0.57$.
The disks now form
stripes
composed of two rows of disks.
For a system with $a=6.0$, $R_{\rm obs}=1.0$, and
$\phi = 0.55$, Fig.~\ref{fig:1}(c) indicates that 
the disks form a density modulated 
state in which stripes appear that are almost four disks wide.
In Fig.~\ref{fig:1}(d),
the disks in a sample with
$a=2.0$, $R_{\rm obs} = 0.5$,
and $\phi = 0.242$
are more uniformly distributed.

\begin{figure}
\includegraphics[width=\columnwidth]{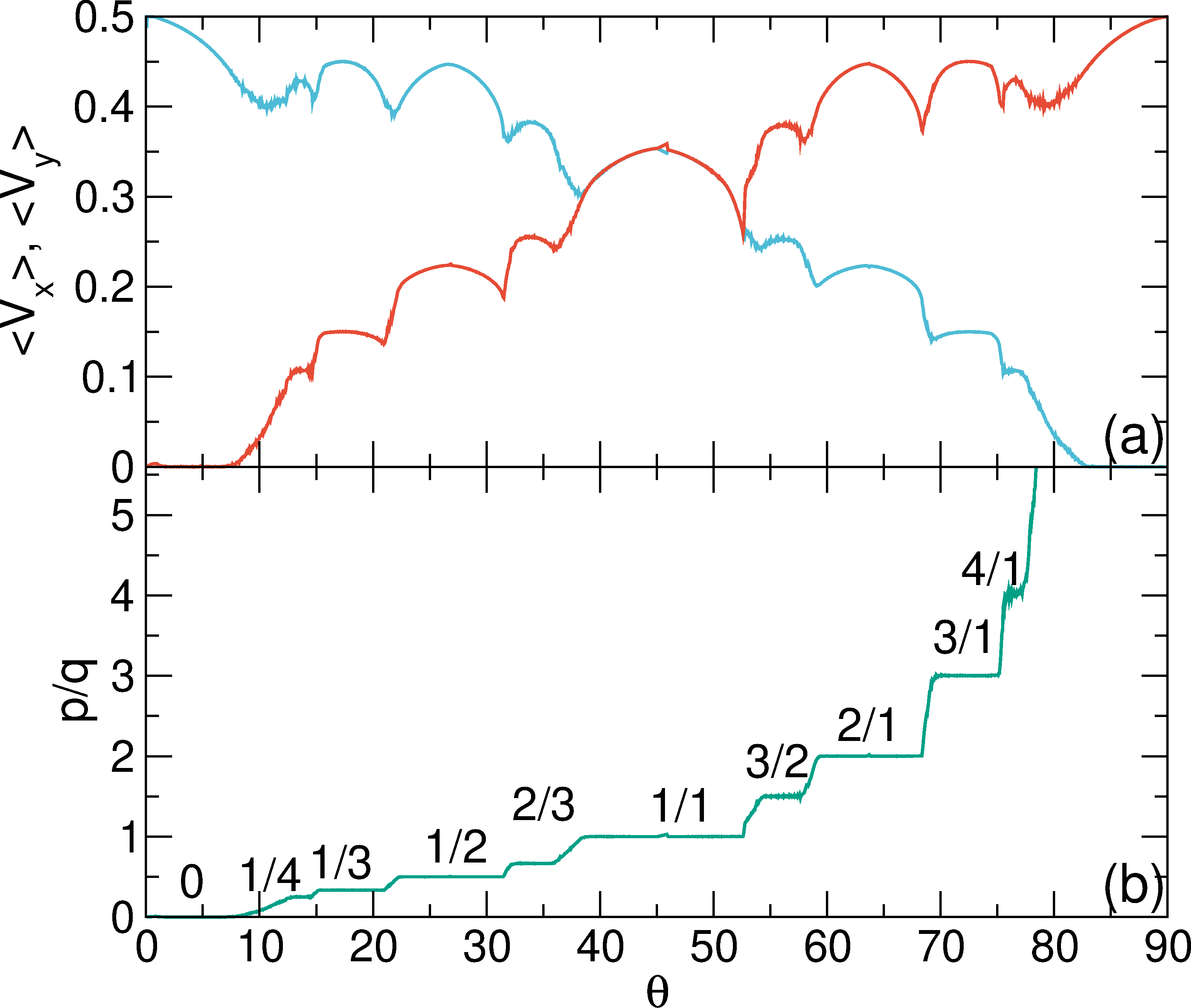}
\caption{
(a) The velocities $\langle V_x\rangle$ in the $x$-direction (blue) and
$\langle V_y\rangle$ in the $y$-direction (red)
vs drive angle $\theta$ for the system in Fig.~\ref{fig:1}(a)
with $a=4.0$, $R_{\rm obs}=0.35$, and $\phi=0.217$.
(b) $p/q$ or $\langle V_{y}\rangle/\langle V_{x}\rangle$
vs $\theta$ showing a series of directional locking steps.
The steps at $0$, 1/4, 1/3, 1/2, 2/3, 1/1, 3/2, 2/1, 3/1, and $4/1$
are highlighted.
}
\label{fig:2}
\end{figure}

In Fig.~\ref{fig:2}(a) we plot
$\langle V_{x}\rangle$ and
$\langle V_{y}\rangle$ versus the angle
$\theta$ of the for the system 
in Fig.~\ref{fig:1}(a) at a drive of $F^D=0.5$,
while in Fig.~\ref{fig:2}(b) we show the corresponding
value of $p/q$ or $\langle V_{y}\rangle/\langle V_{x}\rangle$ versus $\theta$.
The velocity steps in Fig.~\ref{fig:2}(a) are not flat
but
take the form of rounded humps bracketed by cusps at the jumps in and out
of the locking phases.
In contrast, we do find flat steps in the value of $p/q$,
indicating that although the velocity of the disks
is changing, the direction of disk motion is fixed.
On the $1/1$ step,
Fig.~\ref{fig:2}(b) indicates that
the direction of motion is locked over a fixed interval centered
at $\theta = 45^{\circ}$,
while prominent steps
also appear at $p/q = 0$, 1/4, 1/3, 1/2, 2/3, 1, 3/2, 2, 3, and $4$.
The upper edge of the $p/q = 0$ locking step
falls at $\theta = 7^{\circ}$.

\begin{figure}
\includegraphics[width=\columnwidth]{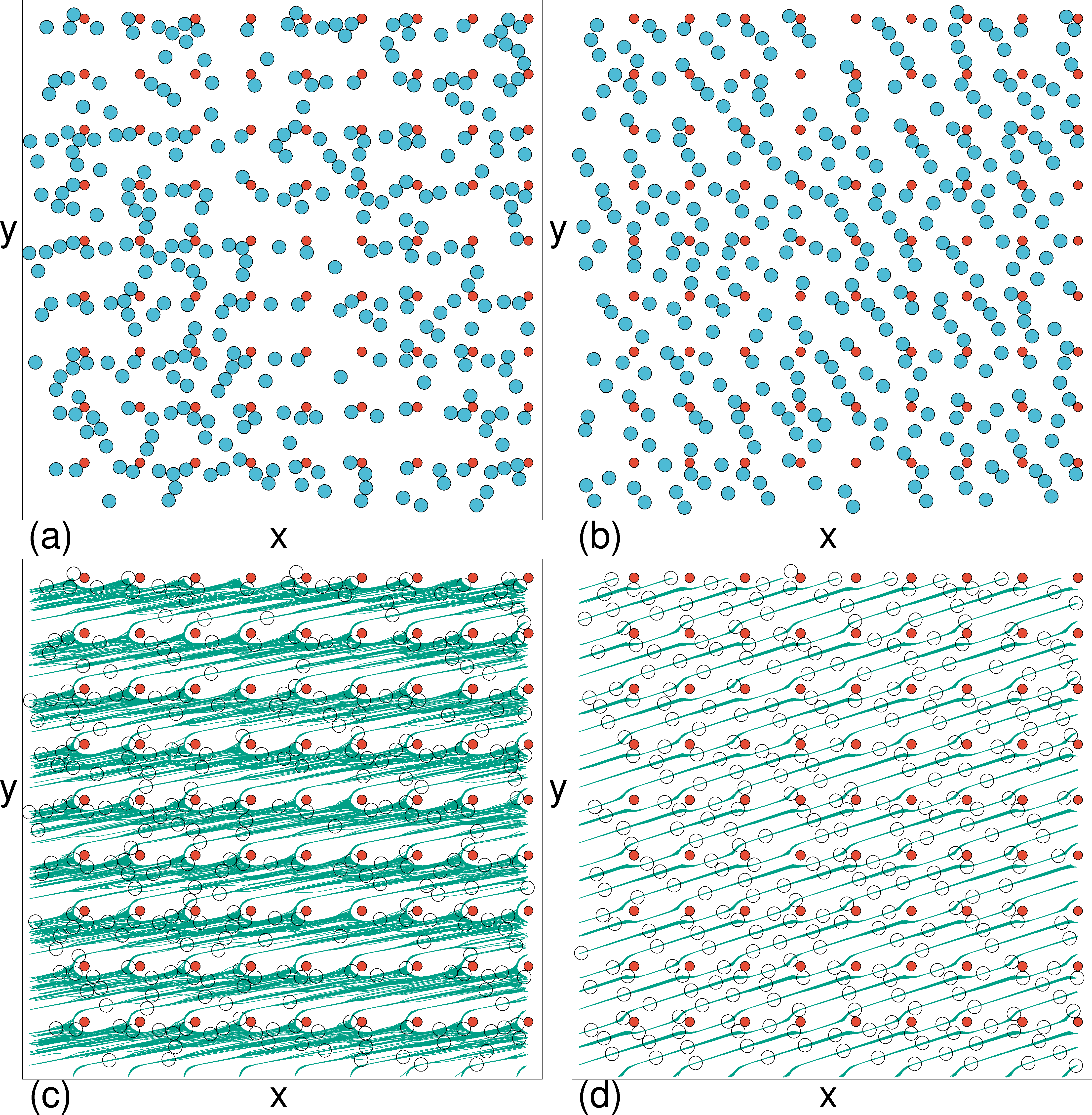}
\caption{(a) Obstacle (red) and disk (blue) locations
for the system in 
Fig.~\ref{fig:2} with $a=4.0$, $R_{\rm obs}=0.35$,
$\phi=0.217$, and $R_d=0.5$.
(a) A non-locking regime
at a drive angle of $\theta=11.5^\circ$, where the disk
configuration is disordered.
(b) The $p/q = 1/3$ step at a drive angle
of $\theta=18.4^\circ$ where the disks are more ordered. 
(c) Disordered disk trajectories (green lines) for the system in
panel (a).
(d) 1D ordered disk trajectories for
the system in panel (b).
In panels (c) and (d), the mobile disks are drawn as open circles
for clarity.}          
\label{fig:3}
\end{figure}

In Fig.~\ref{fig:3}(a)
we show a snapshot of the disks
from the system in Fig.~\ref{fig:2}
at $\theta = 11.5^\circ$ where the motion is not locked and the disk
structure is disordered.
Figure~\ref{fig:3}(b) 
illustrates the
$p/q  = 1/3$ locking step in the same system
at a drive angle of $\theta = 18.4^\circ$,
where
the
disks are more ordered and
have a
weak density modulation
perpendicular to the driving direction.
In Fig.~\ref{fig:3}(c) we
plot the disk trajectories during a fixed time for the sample in
Fig.~\ref{fig:3}(a).
The trajectories are disordered,
with a mixing 
character, and over long times a given disk diffuses gradually
through the sample, indicating that this is a liquid state.
Figure~\ref{fig:3}(d) shows the trajectories for
the $p/q = 1/3$ step from Fig.~\ref{fig:3}(b), where the trajectories    
are strongly ordered and form 1D patterns
oriented along
$\theta_L = \arctan(1/3) = 18.435^{\circ}$.
Here, each disk maintains the same neighbors as it moves,
and there is no long time diffusion.

\begin{figure}
\includegraphics[width=\columnwidth]{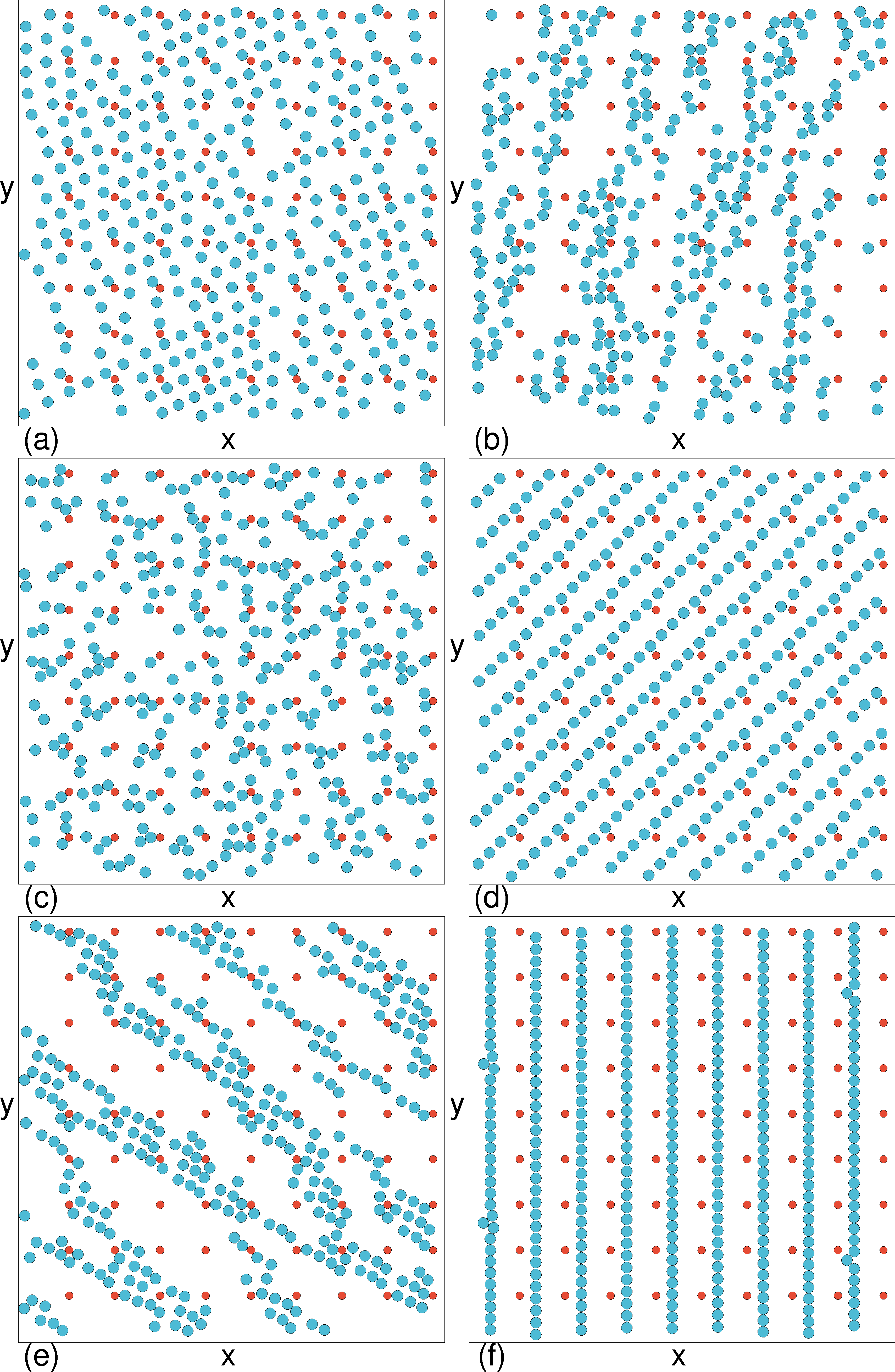}
\caption{
Obstacle (red) and disk (blue) locations for the system in Fig.~\ref{fig:2}
with $a=4.0$, $R_{\rm obs}=0.35$, $\phi=0.217$, and $R_d=0.5$.
(a)
A square lattice configuration at $p/q = 1/2$.
(b) Cluster formation at
$p/q = 2/3$.
(c) A disordered configuration
just below the $p/q = 1/1$ locking step.
(d) Ordered 1D chains at
$p/q = 1/1$.
(e)  A density modulated phase
at $p/q = 3/1$.
(f) Nearly 1D channels at the $90^\circ$ locking step.}    
\label{fig:4}
\end{figure}

In general, the disk motion is more ordered along the locking steps
and more disordered or liquid in the non-locking regimes;
however, within the locking regions, different
structures can arise.
In Fig.~\ref{fig:4}(a) we plot the disk positions for
the system in Fig.~\ref{fig:2} at 
the $p/q = 1/2$ step
where a partially square lattice appears.
On the $p/q=2/3$ step in
Fig.~\ref{fig:4}(b),
the configuration is disordered but shows some
partial clustering or density modulation.
Figure~\ref{fig:4}(c) illustrates the disk positions
in a non-step region just below the
$p/q = 1/1$ step, where the system is disordered but
the density is uniform.
On the $p/q = 1/1$ locking step in
Fig.~\ref{fig:4}(d),
we find ordered 1D chains aligned at $45^\circ$ such that the disks
do not collide with the obstacles.
At $p/q=3/1$ in
Fig.~\ref{fig:4}(e),
a strongly density modulated state appears, while
in Fig.~\ref{fig:4}(f) on the
$90^{\circ}$ locking step,
a series of nearly 1D chains form that are aligned with the $y$-direction.
The 1D lanes on the $90^\circ$ step are more 
ordered than the lanes found
for the $0^\circ$ step in Fig.~\ref{fig:1}(a).
This is the result of the
partial dynamic annealing
produced when the disks pass through
multiple fluctuating non-locked regions before reaching the
$90^\circ$ locking step.
The fluctuating states allow the disks to reach
more ordered configurations,
whereas on the $0^\circ$ locking step,
the disks remain trapped in their initial configuration.
The difference in ordering
between the $0^\circ$ and $90^\circ$ steps would be less pronounced if
the temperature were finite.

\begin{figure}
\includegraphics[width=\columnwidth]{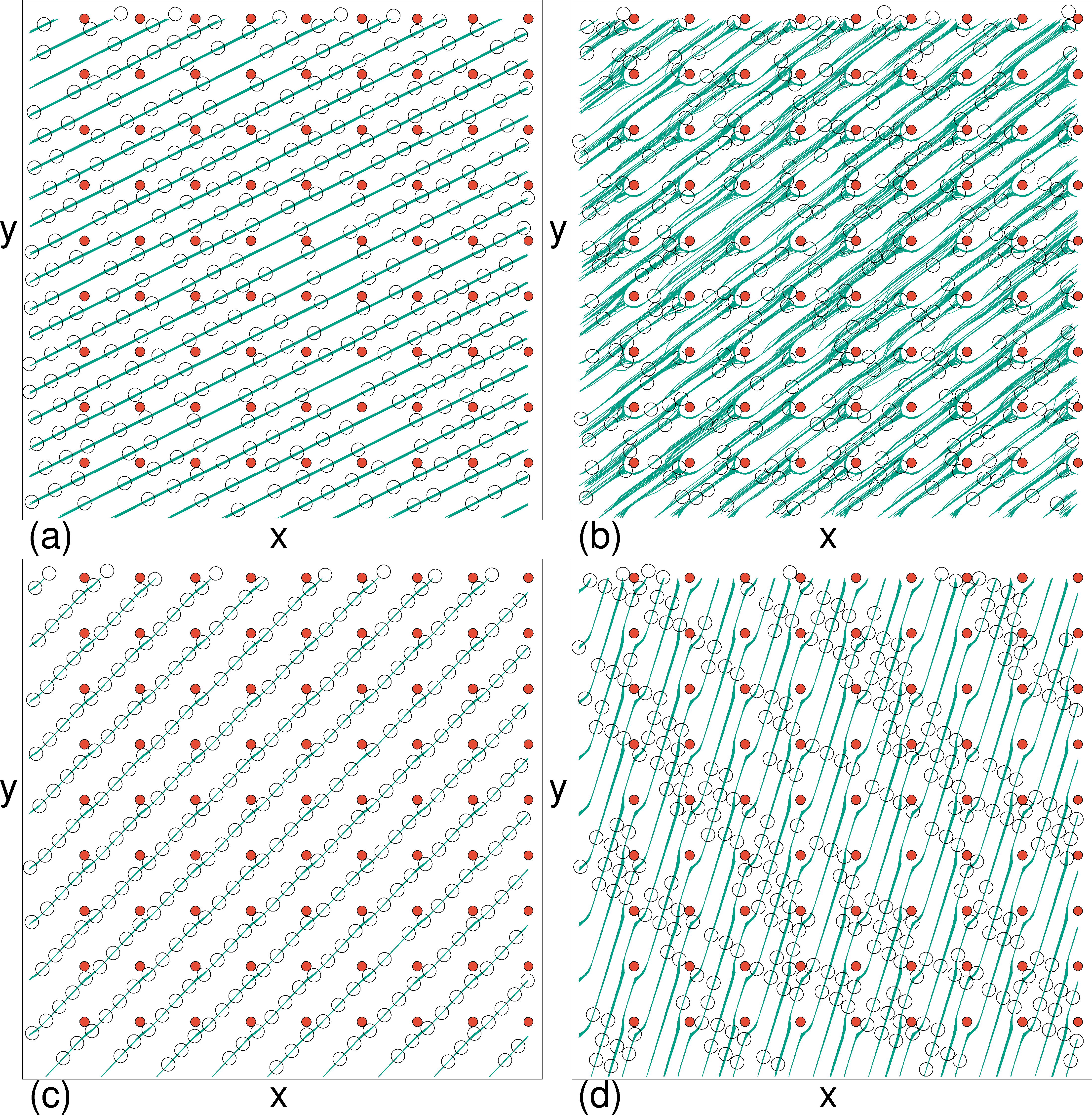}
\caption{
Obstacle (red) and disk (open circle) locations
along with disk trajectories (green)
for the system
in Fig.~\ref{fig:2} with
$a=4.0$, $R_{\rm obs}=0.35$, $\phi=0.217$, and $R_d=0.5$.
(a) 1D trajectories at $p/q =  1/2$.
(b) Disordered trajectories in the nonstep region just below
the $p/q = 1/1$ locking step.
(c) At $p/q = 1/1$, the disks move in 
1D chains along $\theta_L = 45^\circ$.
(d) At $p/q = 3/1$, the system forms a density
modulated phase with ordered trajectories.}

\label{fig:5}
\end{figure}

In Fig.~\ref{fig:5} we plot the disk positions and trajectories
for the system in Fig.~\ref{fig:4}.
At $p/q  = 1/2$ 
in Fig.~\ref{fig:5}(a),
the disks follow 1D paths
and move a distance $2a$ in the $x$ direction for each
translation by $a$ in the $y$ direction.
No collisions occur
between the moving disks and the obstacles.
Figure~\ref{fig:5}(b) shows that the trajectories are disordered
for the system in Fig.~\ref{fig:4}(c)
just below the $p/q = 1/1$ locking step.
In Fig.~\ref{fig:5}(c), the trajectories on the
$p/q = 1/1$ step exhibit ordered 1D motion. 
At $p/q=3/1$ in
Fig.~\ref{fig:5}(d),
the disks form
a disordered density modulated phase in which the trajectories are
ordered.
In general, for the locking phases the
disks move elastically and maintain the same neighbors,
while in the non-locking phases, the disks diffuse with respect
to each other, forming a liquid state.

\begin{figure}
\includegraphics[width=\columnwidth]{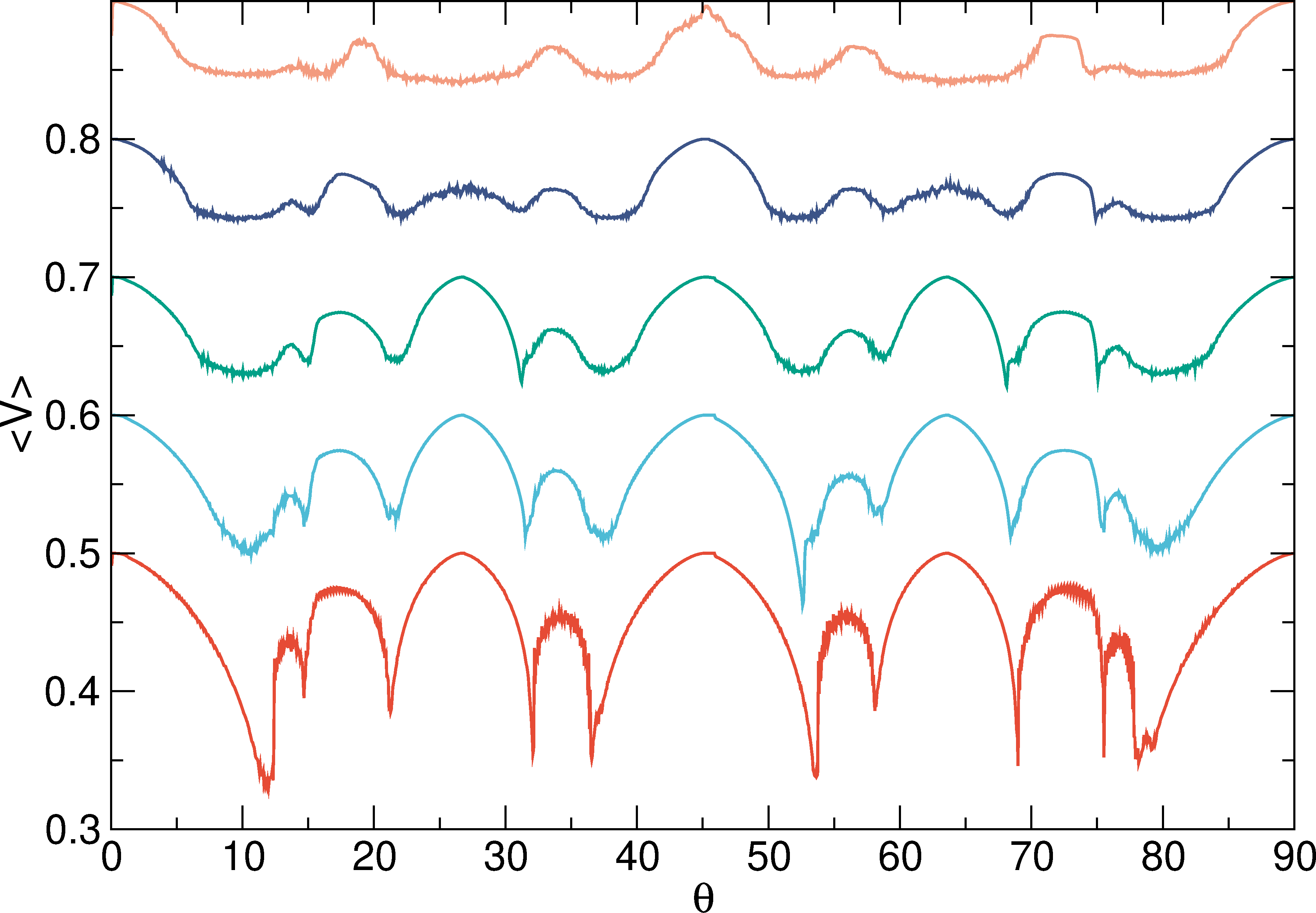}
\caption{ 
The net velocity $\langle V\rangle$ vs $\theta$ for the system in
Fig.~\ref{fig:2} with
$a=4.0$, $R_{\rm obs}=0.35$, and $R_d=0.5$ at
densities
$\phi = 0.096$ (red), $0.217$ (light blue), $0.3857$ (green), 
$0.46$ (dark blue), and $0.558$ (orange), from bottom to top. 
The curves have been shifted
vertically by intervals of $0.1$ for clarity.
}
\label{fig:6}
\end{figure}

In Fig.~\ref{fig:6} we plot the net velocity
$\langle V\rangle = \sqrt{\langle V_x\rangle^2 + \langle V_y\rangle^2}$
versus $\theta$ for the system in Fig.~\ref{fig:2}
at densities of $\phi=0.096$, 0.217, 0.3857, 0.46, and $0.558$.
The net velocity passes through a local maximum at the center of each
locking regime, corresponding to the points
at which the interactions between the
disks and the obstacles are minimized.
For the lowest density of $\phi=0.096$,
we find a range of values
$0.35 < \langle V\rangle < 0.5$.
The maximum velocity cannot exceed the
driving force value,
$F^{D} = 0.5$.
We obtain $\langle V\rangle = F^D$ whenever the moving disks cease
interacting with the obstacles, which occurs for
$p/q = 0$, 1/2, 1/1, and in the $90^\circ$ locking phase.
Here the disks move in 1D chains and do not come into contact with the
obstacles,
as illustrated in Fig.~\ref{fig:5}(a) for $p/q = 1/2$ and
in Fig.~\ref{fig:4}(d) for $p/q = 1/1$. 
Some of the 
directionally locked regimes have a reduced maximum velocity
due to disk-obstacle collisions,
as shown in Fig.~\ref{fig:5}(d) for $p/q = 3/1$
and in Fig.~\ref{fig:3}(d) for $p/q =1/3$.
The lowest velocity values
appear
in the non-locking regimes.
As the density of mobile disks increases,
the number of locked phases diminishes
and the maxima and minima in $\langle V\rangle$ become less distinct.
For example, at the highest density of $\phi=0.58$, we find
only small peaks in $\langle V\rangle$ at $p/q = 1/3$, 2/3, 1/1, 3/2,
and $3/1$, while the peaks 
at $1/4$, 1/2, 2/1, and $4/1$ have disappeared.

\begin{figure}
\includegraphics[width=\columnwidth]{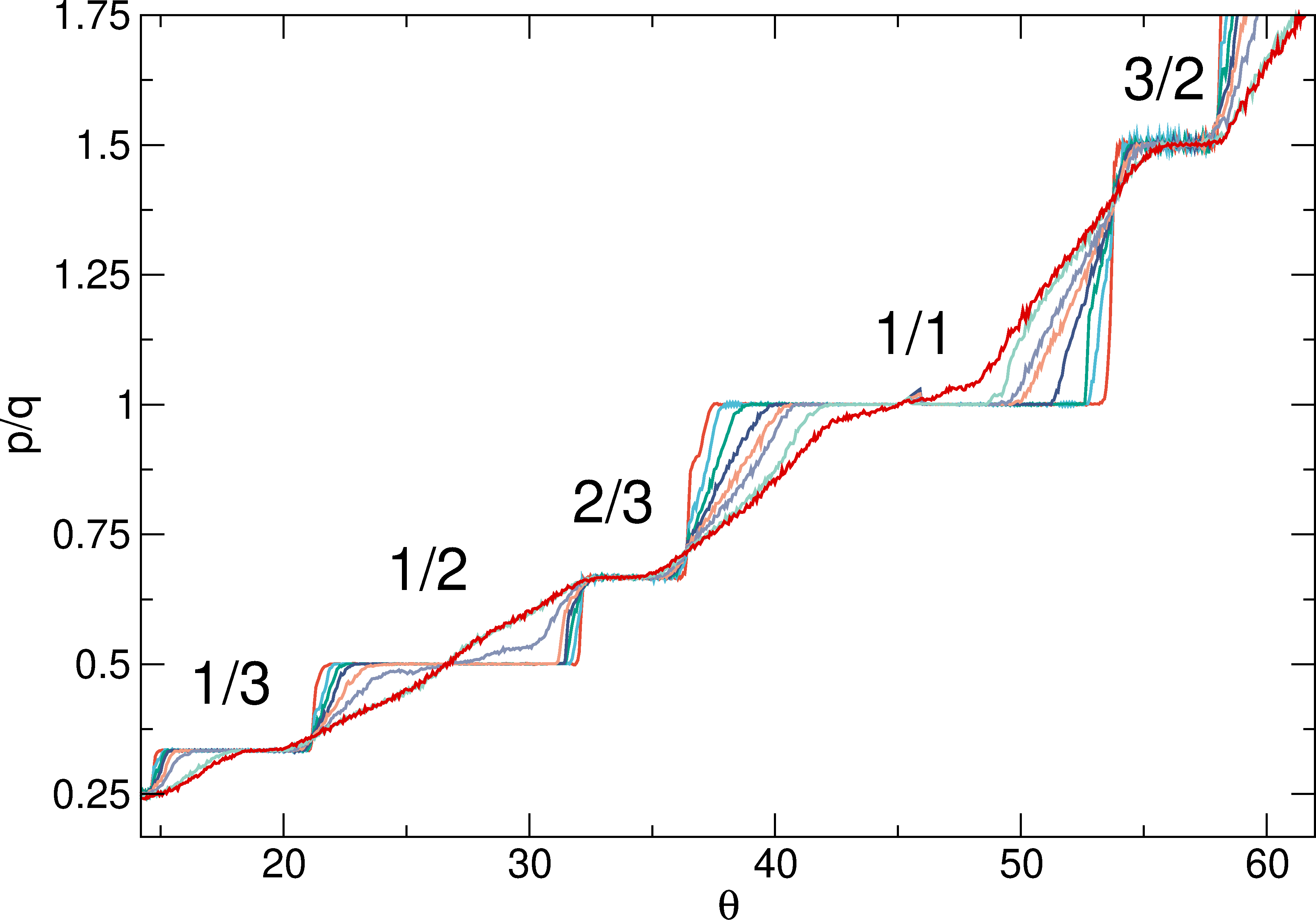}
\caption{
$p/q$ vs $\theta$ for the system in
Fig.~\ref{fig:6} with
$a=4.0$, $R_{\rm obs}=0.35$, and $R_d=0.5$ at
$\phi=0.096$ (dark orange),
0.1567 (light blue),
0.217 (dark green),
0.278 (dark blue),
0.3857 (light orange),
0.46 (light purple),
0.52 (light green),
and 0.558 (red).
The steps gradually disappear as the disk density
increases.
}
\label{fig:7}
\end{figure}

The evolution of the locking regimes is illustrated in the plot of
$p/q$ versus $\theta$
at different values of $\phi$ in Fig.~\ref{fig:7}.
The $p/q=1/3$, 2/3, and $3/2$ steps decrease in width as $\phi$ increases
but remain present even for the highest disk densities, whereas
the $p/q = 1/2$
locking step disappears when
$\phi > 0.5$.
At $\phi = 0.46$,
the 1/2 step is partially locked and
$p/q$ does not remain constant on the step
but shows a linear increase.
For the $p/q = 1/1$ step,
complete locking is lost when
$\phi > 0.53$ and there is only partial locking 
at $\phi = 0.558$. 

\begin{figure}
\includegraphics[width=\columnwidth]{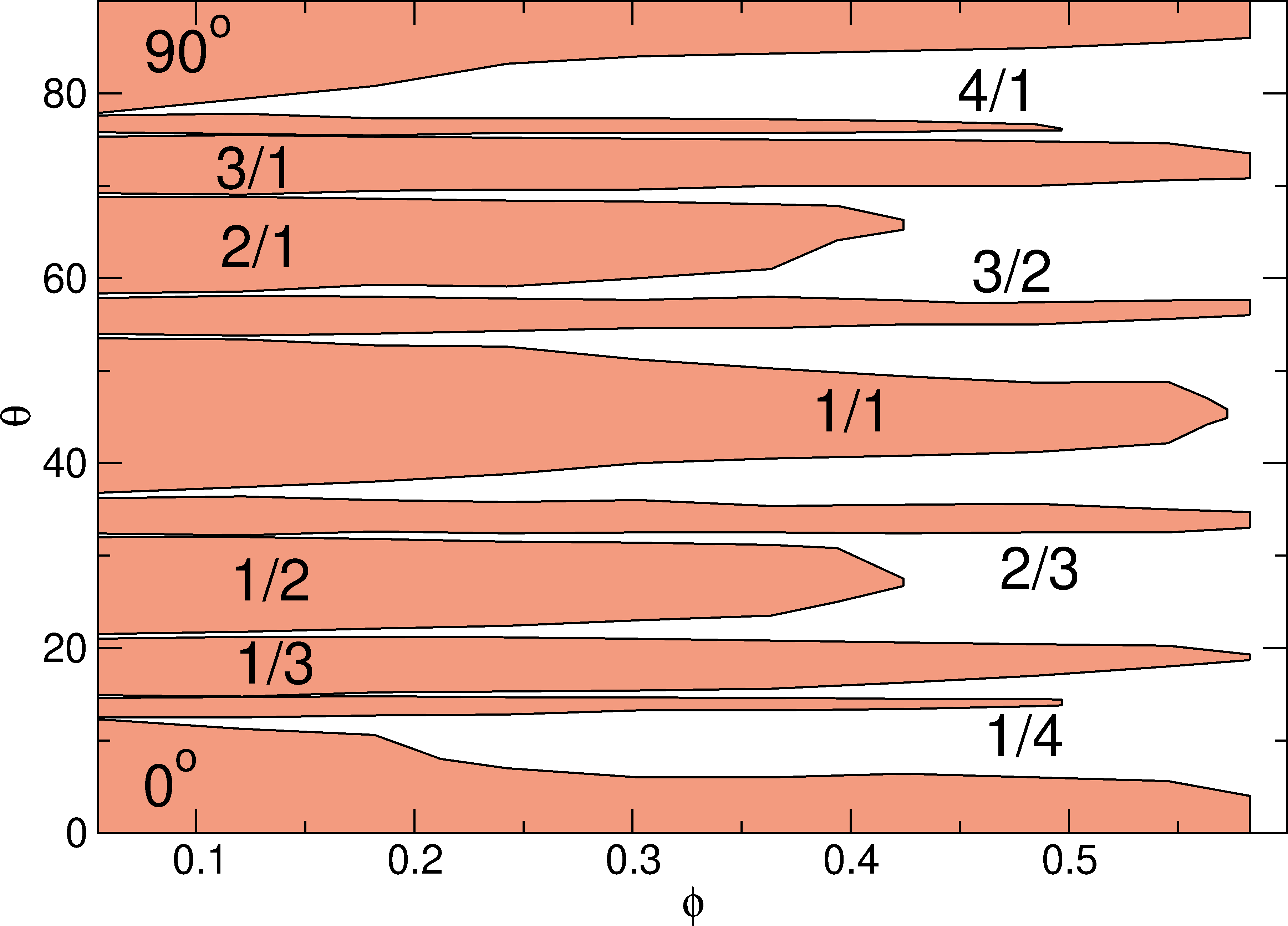}
\caption{
Locked regions (shaded areas) as a function of $\theta$ vs disk density
$\phi$
for the system in Fig.~\ref{fig:7}
with
$a=4.0$, $R_{\rm obs}=0.35$, and $R_d=0.5$.
The $p/q=0$, 1/4, 1/3, 1/2, 2/3, 1/1, 3/2, 2/1, 3/1, 4/1,
and $90^\circ$ steps are labeled.
The $p/q=1/2$ and $2/1$ steps are the first to disappear as $\phi$ increases.
}
\label{fig:8}
\end{figure}

Based on curves such as those shown 
in Fig.~\ref{fig:7},
we construct a phase diagram highlighting the different locking phases.
In Fig.~\ref{fig:8} we plot the locations of the locked phases
as a function of $\theta$ versus $\phi$. 
The most prominent locking steps appear at
$p/q = 1/1$, $0^\circ$, and $90^\circ$. 
The $p/q=1/2$ and $2/1$ steps disappear when
$\phi > 0.4$ and the disks become dense enough that
1D ordered chains can no longer form.
In general, the widths of all of the steps decrease with increasing $\phi$.  

\begin{figure}
\includegraphics[width=\columnwidth]{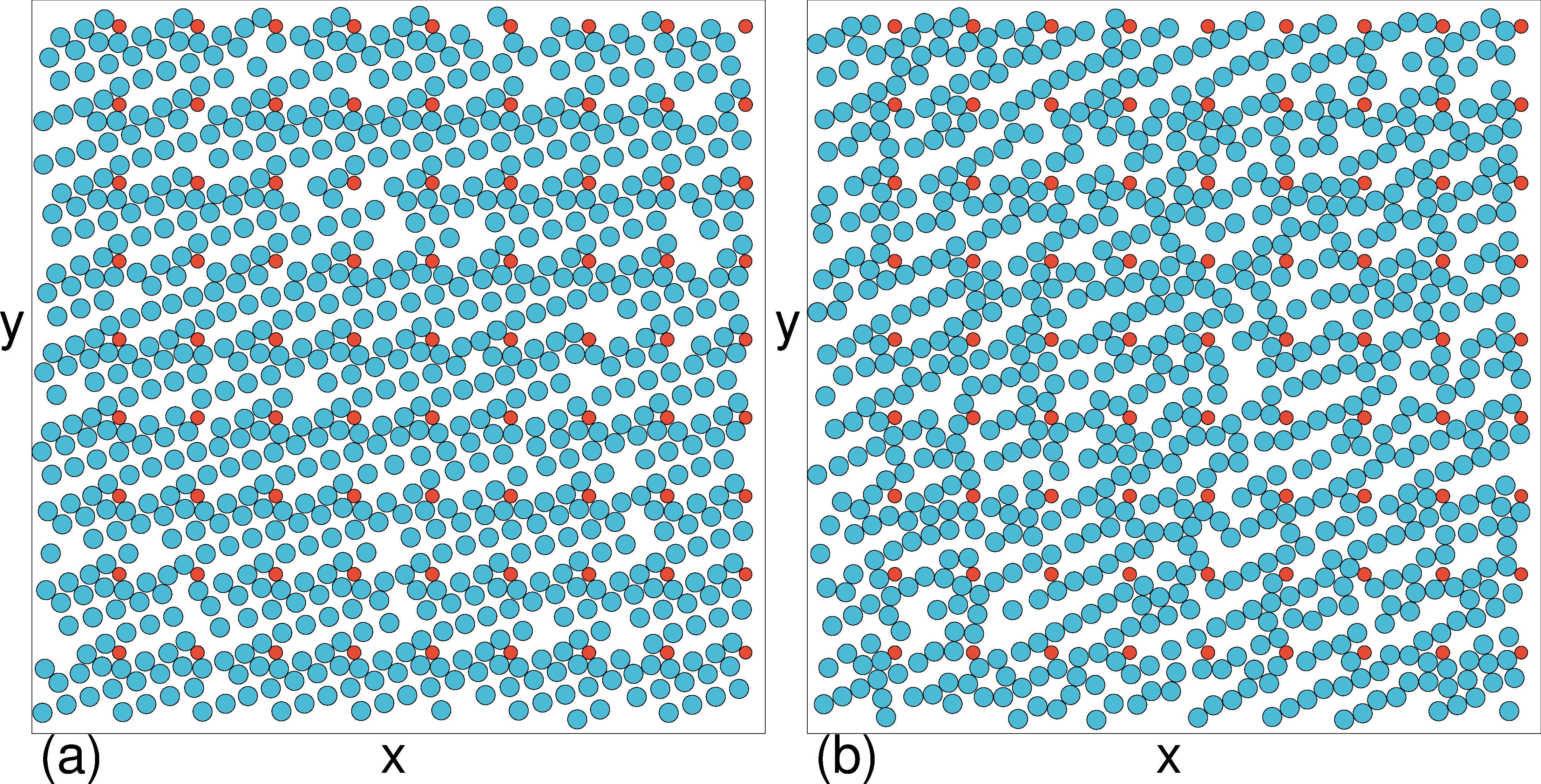}
\caption{(a) Obstacle (red) and disk (blue) locations for
the system in Fig.~\ref{fig:8} with
$a=4.0$, $R_{\rm obs}=0.35$, and $R_d=0.5$
at $\phi=0.52$.
(a)
At $p/q =1/3$,
there is a locking step and the disk positions are partially disordered.
(b)
At $p/q = 1/2$, there is no directional locking
and the disk configuration is disordered.       
}
\label{fig:9}
\end{figure}

In Fig.~\ref{fig:9}(a) we illustrate the disk configurations 
for the system in Fig.~\ref{fig:8}
at $\phi = 0.52$ and $p/q = 1/3$,
where
a directional locking step occurs and
the disks are partially ordered.
In the same system at $p/q = 1/2$,
Fig.~\ref{fig:9}(b) shows that
the disks are much more disordered and there is no
directional locking.     
In general, as the disk density increases,
it is more difficult for the disks to move
around the obstacles in an ordered fashion in order to form a
locked state.

\begin{figure}
\includegraphics[width=\columnwidth]{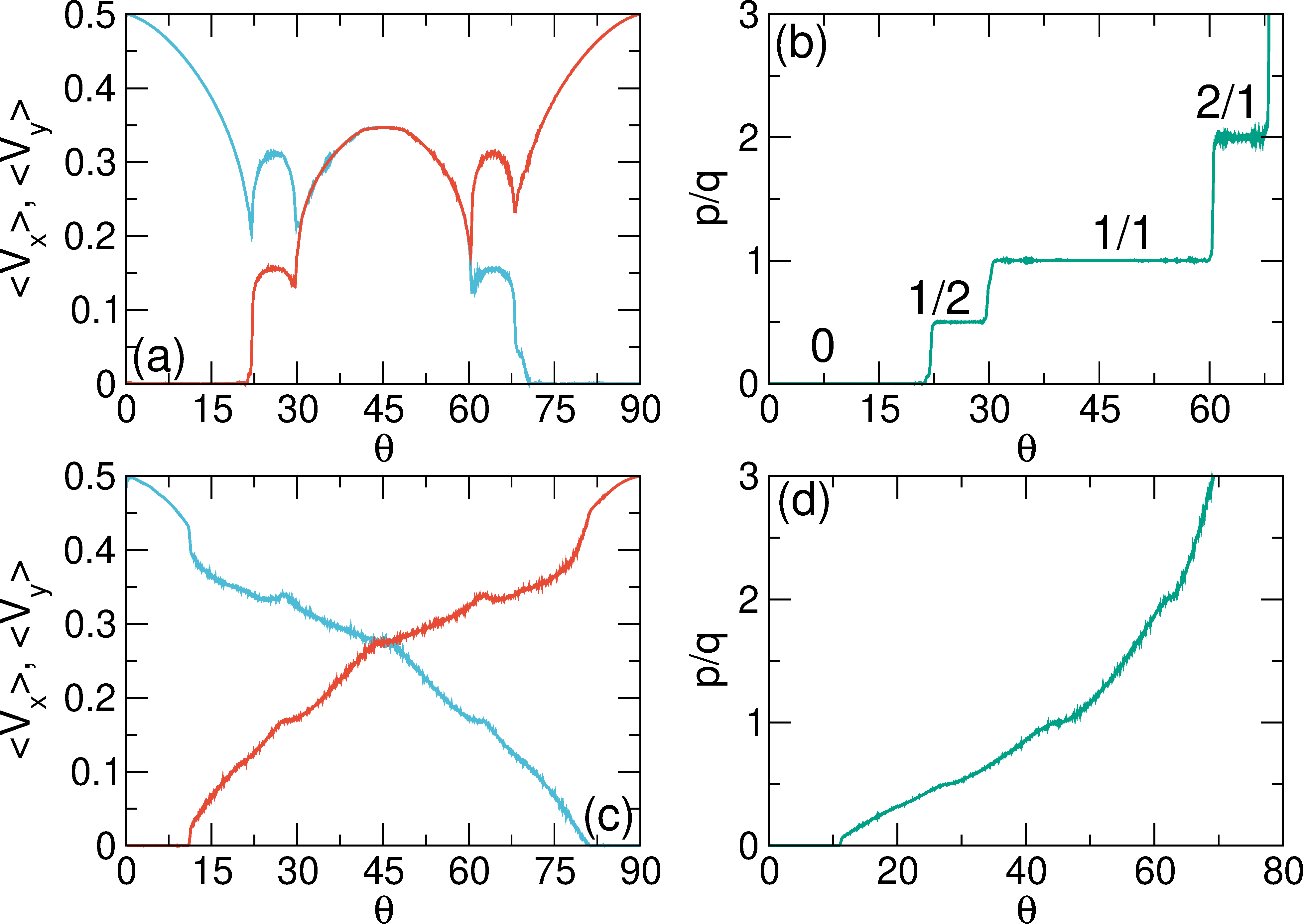}
\caption{(a)
$\langle V_{x}\rangle$ (blue) and $\langle V_{y}\rangle$ (red)
vs $\theta$ for the system in
Fig.~\ref{fig:1}
with $a=4.0$, 
$R_{\rm obs} = 1.0$,
$R_{d} = 0.5$, and $\phi = 0.39$.
(b) The corresponding $p/q$ vs $\theta$ 
showing locking steps at $p/q = 0$, 1/2, 1/1, 2/1, and $90^\circ$. 
(c) $\langle V_{x}\rangle$ (blue) and $\langle V_{y}\rangle$ (red)
vs $\theta$ for the system in 
Fig.~\ref{fig:1}(b) with
$\phi = 0.57$, where locking steps appear
only at $0^\circ$ and $90^\circ$.  (d) 
The corresponding $p/q$ vs $\theta$.
}
\label{fig:10}
\end{figure}

The width of the locking steps as a function of $\phi$
is affected by the value of the obstacle radius $R_{\rm obs}$.
In Fig.~\ref{fig:10}(a) we plot $\langle V_{x}\rangle$ and
$\langle V_{y}\rangle$ versus $\theta$ for
the system from Fig.~\ref{fig:1} with
$R_{\rm obs} = 1.0$, $R_{d} = 0.5$, and $\phi = 0.39$,
while in Fig.~\ref{fig:10}(b) we show
the corresponding $p/q$ versus $\theta$.
Here, locking steps appear only for $p/q = 0$, 1/2, 1/1, 2/1, and
$90^\circ$.
Figures~\ref{fig:10}(c,d) show
$\langle V_{x}\rangle$, $\langle V_{y}\rangle$,
and $p/q$ versus $\theta$
for the same system at a higher mobile disk density 
of $\phi = 0.57$.
Locking occurs
only at angles of $0^\circ$ and $90^\circ$,
while the other locking steps are lost.

\begin{figure}
\includegraphics[width=\columnwidth]{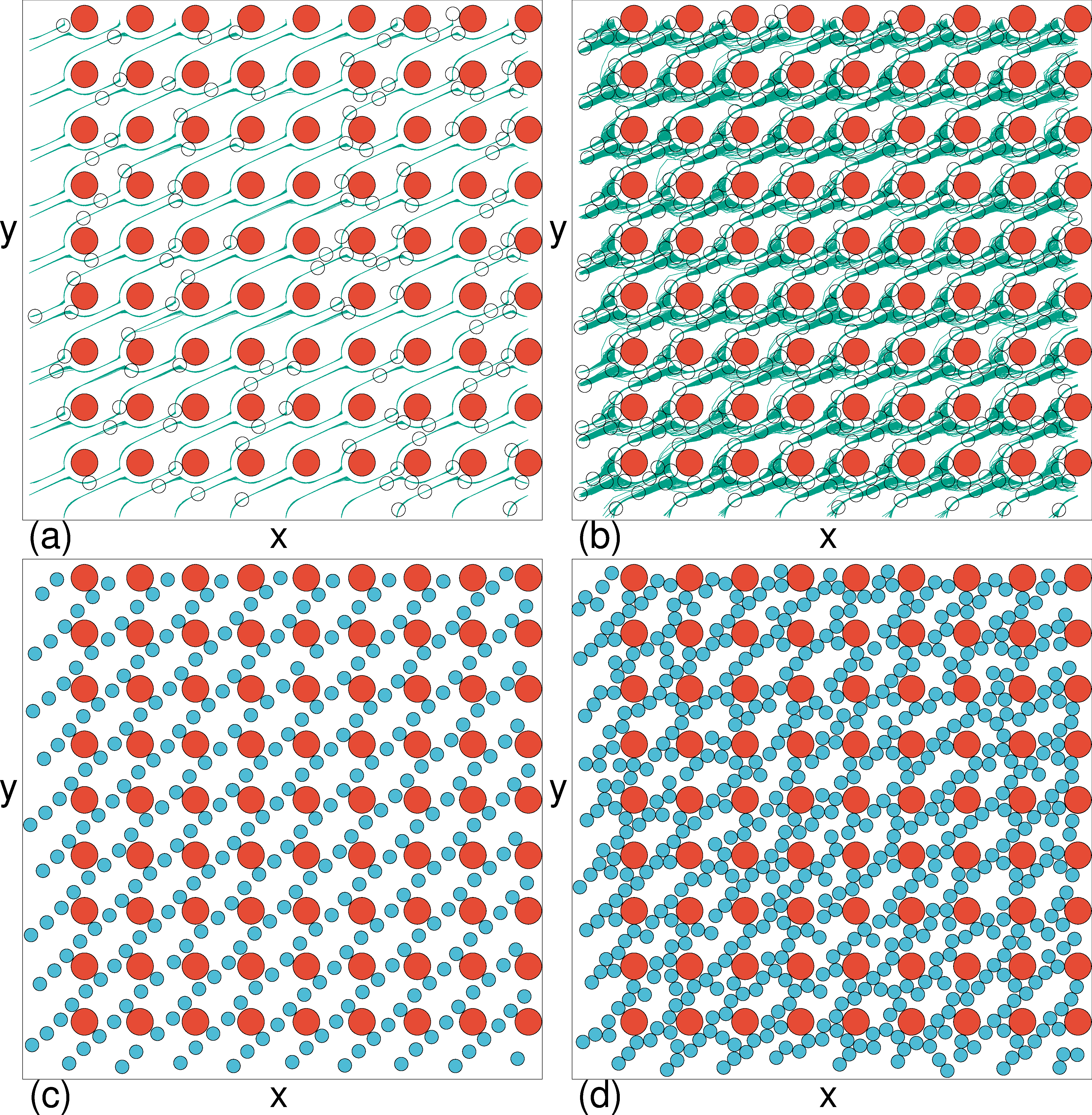}
\caption{
(a) Obstacle (red) and disk (blue or open circle) locations along with
disk trajectories (green)
for the system in Fig.~\ref{fig:10}(a)
with
$a=4.0$, $R_{\rm obs}=1.0$, $R_d=0.5$, and $\phi=0.39$ in
the $p/q = 1/2$ locking phase where the disk
configurations are heterogeneous but the trajectories are ordered.
(b) The
same system at a higher
mobile disk density of $\phi = 0.51$
in the nonlocking regime just above the $p/q = 1/2$ locking 
step, where the trajectories are disordered.
(c) Image without trajectories
of the $p/q = 1/1$ locking step
for the same system at
$\phi=0.46$
and $\theta = 45^\circ$.
(d) The system in Fig.~\ref{fig:10}(c) with $\phi=0.57$
at $p/q=1/1$,
where there is no directional locking
and the disk
configuration is disordered.      
}
\label{fig:11}
\end{figure}

In Fig.~\ref{fig:11}(a)
we show the disk configurations and trajectories 
for the system in
Fig.~\ref{fig:10}(a) on the $p/q = 1/2$ locking step
where the system forms 
a partially clustered state with ordered trajectories.
Figure~\ref{fig:11}(b) illustrates the
disordered disk trajectories that appear in
the same system at $\phi = 0.51$
for the non-locking regime just above the $p/q = 1/2$ locking step.
In Fig.~\ref{fig:11}(c) we plot the disk positions without trajectories
for a sample with $\phi=0.46$ at $\theta = 45^\circ$
on the $p/q =1/1$ locking step,
where the disks move in 1D channels
oriented $45^\circ$ from the $x$ axis.
The same system is shown at $\phi=0.57$
in Fig.~\ref{fig:11}(d),
where the locking step is absent and the disk configuration is disordered.
As $\phi$ increases,
ordered locking flow can occur at $p/q = 1/1$
as long as the chains of moving disks remain narrow enough
to avoid colliding with the obstacles while moving.
When the obstacle radius is
small,
the number $n$ of rows of moving disks that can fit between the obstacles
is limited by the diameter of the disks.
Specifically, if $n$ rows of disks are driven at an angle $\theta$,
obstacle-disk collisions can be avoided only if
$nR_d + R_{\rm obs}\leq \frac{a}{2}\cos{\theta}$,
giving the criterion
$n \leq (1/R_d)(\frac{a}{2}\cos{\theta}-R_{\rm obs})$.
For
the system in Fig.~\ref{fig:11} with
$a = 4.0$, $R_{d} = 0.5$, and $R_{\rm obs} = 1.0$, we obtain
$n \leq 2$ when $\theta=0$, indicating that locking of two rows
can occur on the $p/q=0$ step, while
$n \leq 0.8$ when $\theta=45^\circ$, showing that no locking occurs
at $p/q=1/1$.

\begin{figure}
\includegraphics[width=\columnwidth]{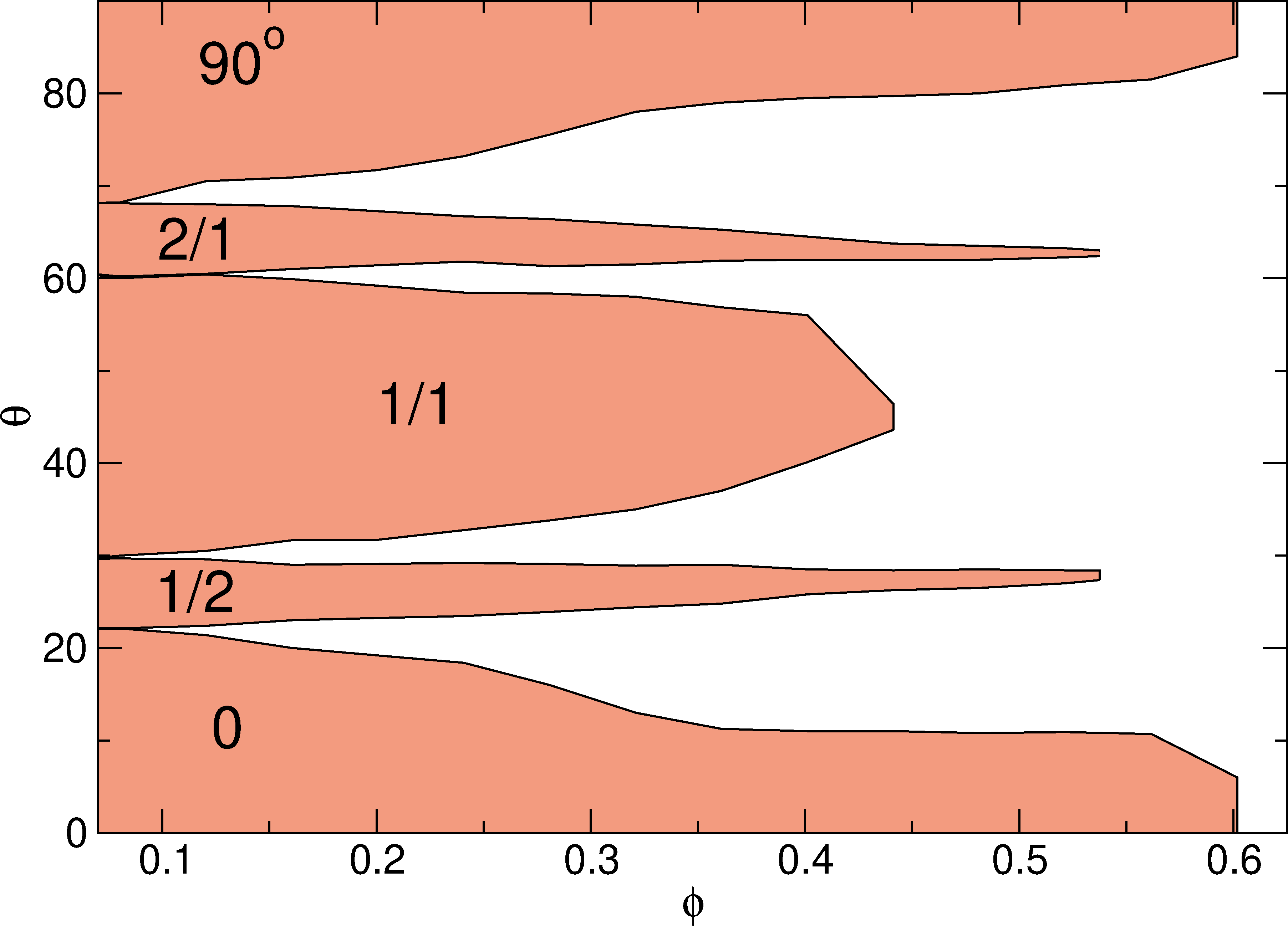}
\caption{
Locked regions (shaded areas) as a function
of $\theta$ vs $\phi$ for the system in Fig.~\ref{fig:10}
with $a=4.0$, $F^D=0.5$,
$R_{\rm obs} = 1.0$,
and $R_{d} = 0.5$,
showing directional locking
at $p/q=0$, 1/2, 1/1, 2/1, and $90^\circ$.
The $1/1$ step is lost when $\phi > 0.4$.     
}
\label{fig:12}
\end{figure}

In Fig.~\ref{fig:12} we show the locations of the locking steps as a function
of $\theta$ versus $\phi$ for the system in
Fig.~\ref{fig:10}.
There are only five steps at
$p/q=0$, 1/2, 1/1, 2/1, and $90^\circ$.
The $1/1$ step disappears when
$\phi > 0.45$,
the $1/2$ and $2/1$ steps vanish above
$\phi=0.55$, and the $0^\circ$ and $90^\circ$ steps persist up to
$\phi=0.6$.
As $R_{\rm obs}$ increases,
the number of
locking phases decreases.

\section{Varied Obstacle Radius and Clogging}

\begin{figure}
\includegraphics[width=\columnwidth]{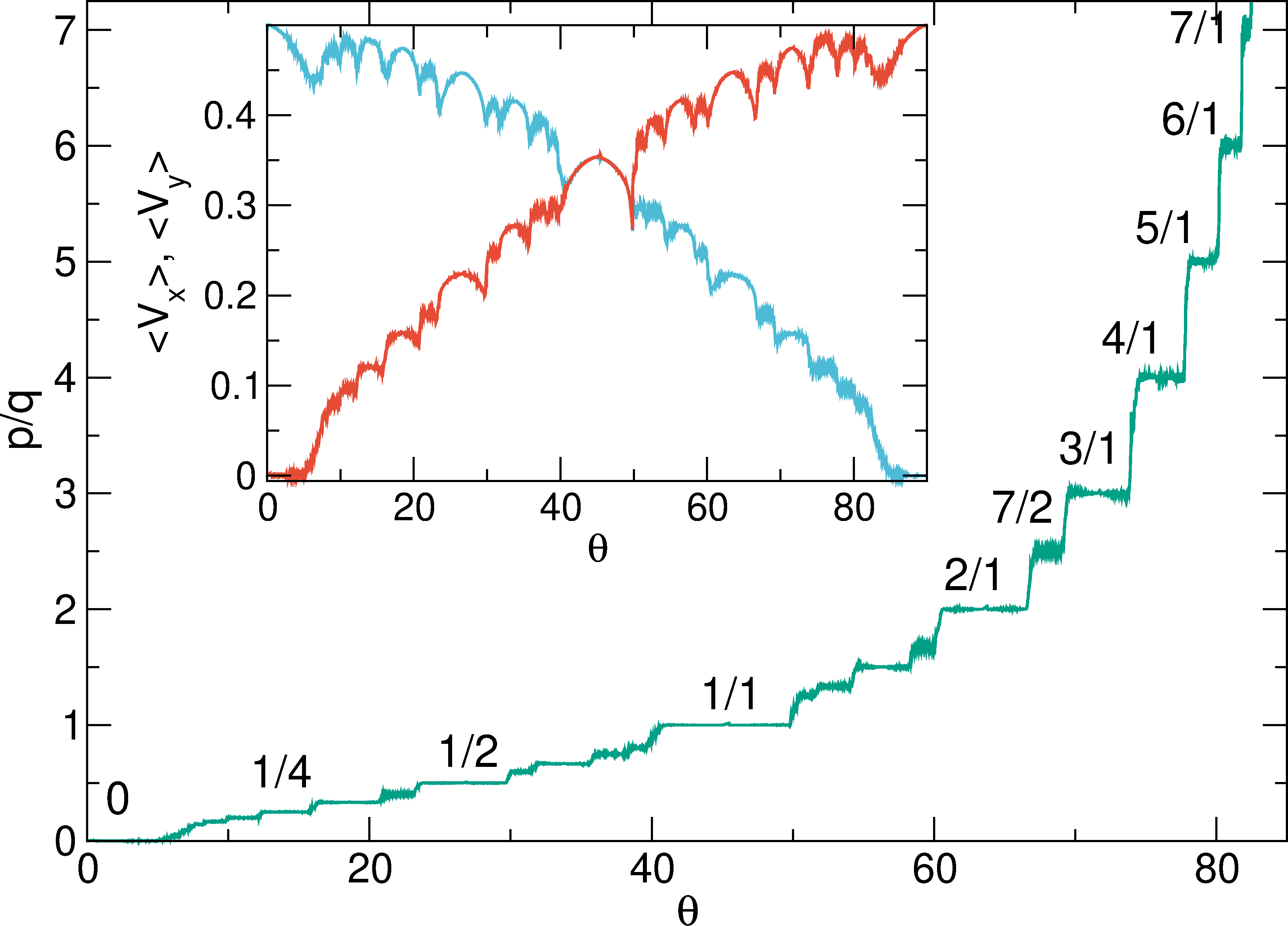}
\caption{
Inset: $\langle V_{x}\rangle$ (blue) and $\langle V_{y}\rangle$ (red)
vs $\theta$ for a system with $R_{d} = 0.5$, $R_{\rm obs} = 0.025$, $a=4.0$, and
$\phi = 0.1934$.
Main panel: $p/q$ vs $\theta$ for the same system over the range
$0 \leq \theta \leq 86^{\circ}$.
Locking steps with
$p/q  = 0$, 1/4, 1/1, 2/1, 7/2, 3/1, 4/1, 5/1, 6/1, and $7/1$
are labeled. There is also another
step at $p/q=8/1$ which is not shown.  
}
\label{fig:13}
\end{figure}

We next vary the radius of the obstacles while holding the
lattice constant at $a=4.0$ and fixing the number of mobile disks.
As the obstacle radius decreases,
we find a larger number of possible locking phases.
In the inset of Fig.~\ref{fig:13} we plot 
$\langle V_{x}\rangle$ and $\langle V_{y}\rangle$
versus $\theta$ for a system with $R_{d} = 0.5$, $R_{\rm obs} = 0.025$, and 
$\phi = 0.1934$.
Here $N_{\rm obs}=81$ and $N_d=319$.
A series of dips appear in the velocity curves at the edges of each
locking phase, and on the $p/q=1/1$ locking step we find
$\langle V_x\rangle=\langle V_y\rangle$.
In the main panel of Fig.~\ref{fig:13}(b),
we show the corresponding $p/q$ versus $\theta$
curve up to $\theta = 86^{\circ}$,
with labels indicating the steps where
$p/q  = 0$, 1/4, 1/1, 2/1, 7/2, 3/1, 4/1, 5/1, 6/1, and $7/1$.
There is an additional
step at $p/q=8/1$ which is not shown.

\begin{figure}
\includegraphics[width=\columnwidth]{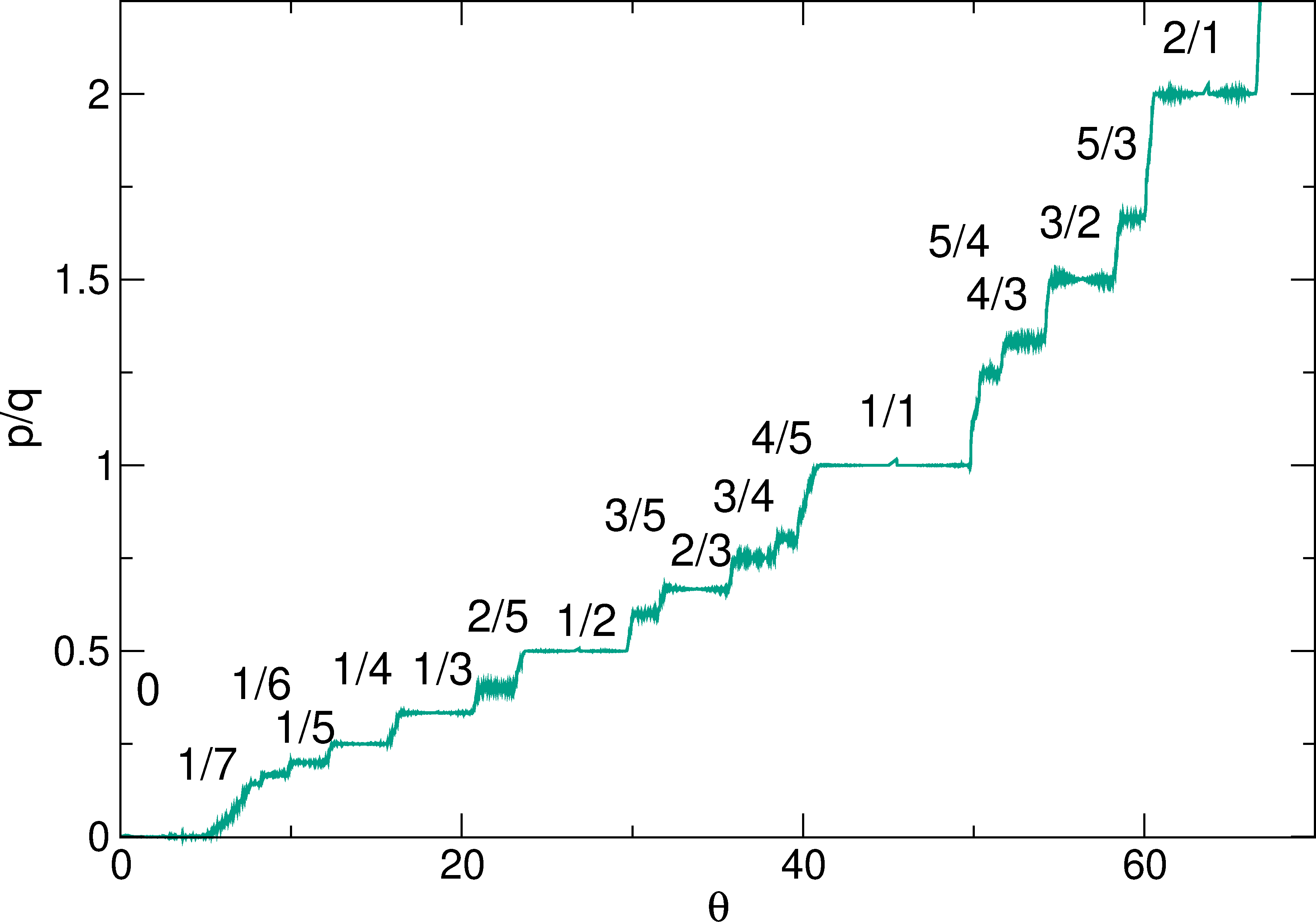}
\caption{
A zoomed in view of
$p/q=\langle V_y\rangle/\langle V_x\rangle$ vs $\theta$
for the system from Fig.~\ref{fig:13}
with 
$R_{d} = 0.5$, $R_{\rm obs} = 0.025$, $a=4.0$, and
$\phi = 0.1934$.
Labels indicate the locking steps at $p/q = 0$,
1/7, 1/6, 1/5, 1/4, 1/3, 2/5, 1/2, 3/5, 2/3,
3/4, 4/5, 1/1, 5/4, 4/3, 3/2, 5/3, and $2/1$. 
}
\label{fig:14}
\end{figure}

In Fig.~\ref{fig:14} we show a blowup of the $p/q$ versus $\theta$ curve 
from the main panel of Fig.~\ref{fig:13}(b)
to better illustrate the additional 
locking phases at $p/q = 0$, 1/7, 1/6, 1/5, 1/4, 1/3, 2/5, 1/2, 3/5, 2/3,
3/4, 4/5, 1/1, 5/4, 4/3, 3/2, 5/3, and $2/1$.
As the
obstacle density
decreases, the number of possible
$p/q$ locking steps increases, as described in Section V.
The presence of the higher order
values of $p/q$ is limited  by the radius 
and density of the mobile disks as well as by the system size.
For example, if we
increase the number of mobile disks so that $\phi$ is larger,
the higher order locking steps disappear.

\begin{figure}
\includegraphics[width=\columnwidth]{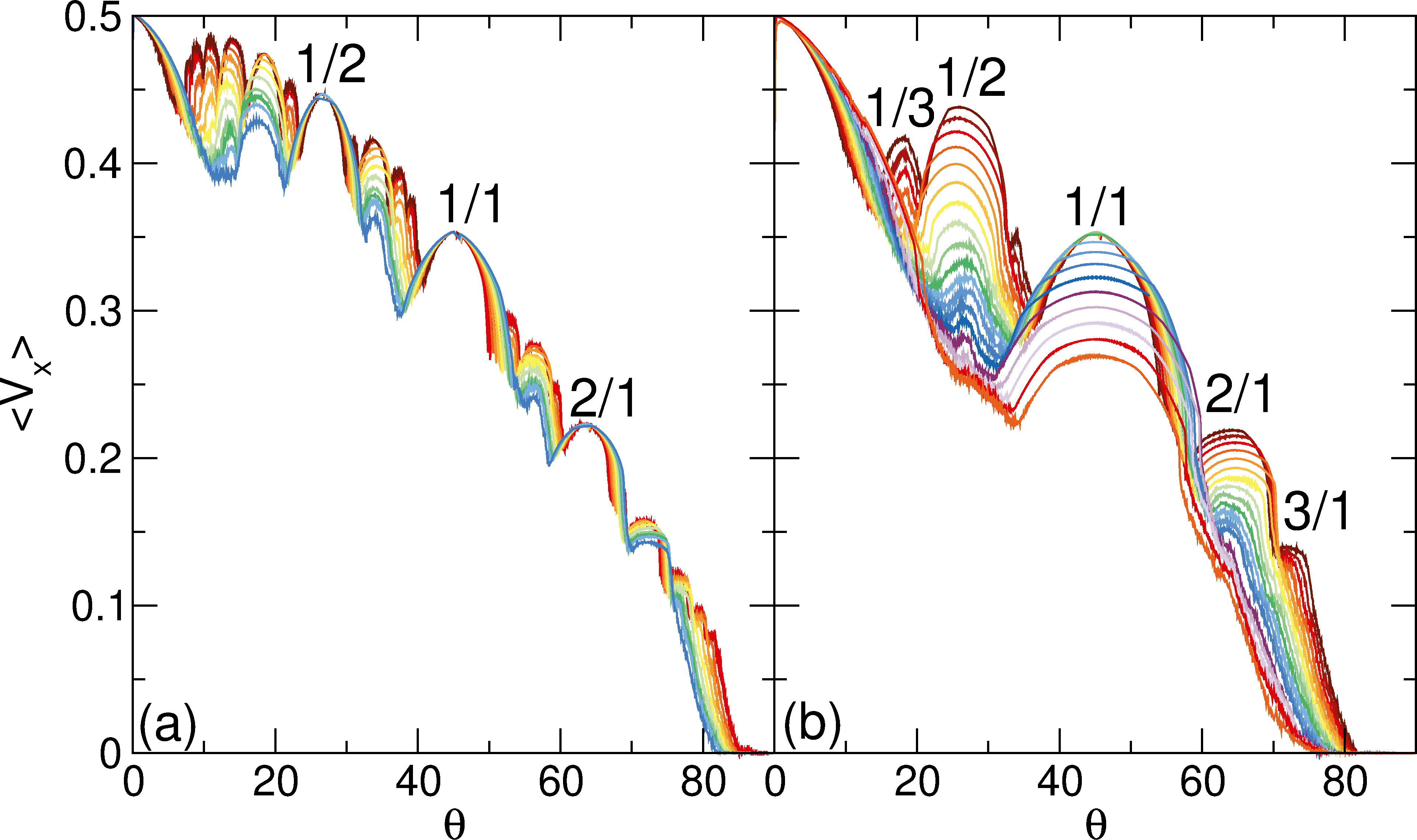}
\caption{
(a) $\langle V_{x}\rangle$ vs $\theta$ curve for the system in
Fig.~\ref{fig:13}
with $R_d=0.5$, $a=4.0$, and $\phi=0.1934$
at $R_{\rm obs} = 0.0125$,
0.025, 0.05, 0.1, 0.15, 0.2, 0.25, 0.3, 0.35, 0.4, and $0.45$,
from top to bottom.
(b)
The same
for $R_{\rm obs}$
ranging from $R_{\rm obs}= 0.05$ to $R_{\rm obs}=1.45$ in
intervals of $0.05$, from top to bottom.
As $R_{\rm obs}$ increases, both the velocity
and the number of steps decrease.}
\label{fig:15}
\end{figure}

In Fig.~\ref{fig:15}(a) we plot
$\langle V_{x}\rangle$ versus $\theta$ for the system in
Fig.~\ref{fig:13} for varied
$R_{\rm obs}$ from $R_{\rm obs}=0.0125$ to $R_{\rm obs}=0.45$.
At the $p/q = 1/2$, 1/1, and $2/1$ locking steps,
$\langle V_{x}\rangle$ reaches the same maximum
value regardless of the value of $R_{\rm obs}$.
At the maximum of each of these steps,
the disks no longer collide with the obstacles
so the velocity is insensitive to the obstacle radius.
For other locking steps,
there are always disk-obstacle collisions and therefore
$\langle V_x\rangle$ decreases with increasing $R_{\rm obs}$.
In general, for each $p/q$ step there is a particular value 
of $R_{\rm obs}$ above which 
$\langle V_{x}\rangle$ begins to decrease
with increasing $R_{\rm obs}$.
The higher order $p/q$ locking phases
also gradually disappear as
$R_{\rm obs}$ increases.

In Fig.~\ref{fig:15}(b) we plot $\langle V_x\rangle$ versus $\theta$ for
the same system as in Fig.~\ref{fig:15}(a)
over the range $0.5 \leq R_{\rm obs}\leq 1.45$.
At the peak of the $p/q=1/1$ locking step,
the value of $\langle V_x\rangle$ remains constant
for $R_{\rm obs} \leq 0.9$, while for larger
values of $R_{\rm obs}$, $\langle V_x\rangle$ begins to decrease. 
The width of the
$p/q=1/1$ step increases with increasing $R_{\rm obs}$ 
up to $R_{\rm obs} = 1.35$, after which it decreases again,
while the steps with
$p/q = 1/3$, 1/2, 2/1, and $3/1$
decrease in width until for
$R_{\rm obs} = 1.45$ they are absorbed by the
$0^\circ$, $90^\circ$, and $p/q=1/1$ locking regimes.

\begin{figure}
\includegraphics[width=\columnwidth]{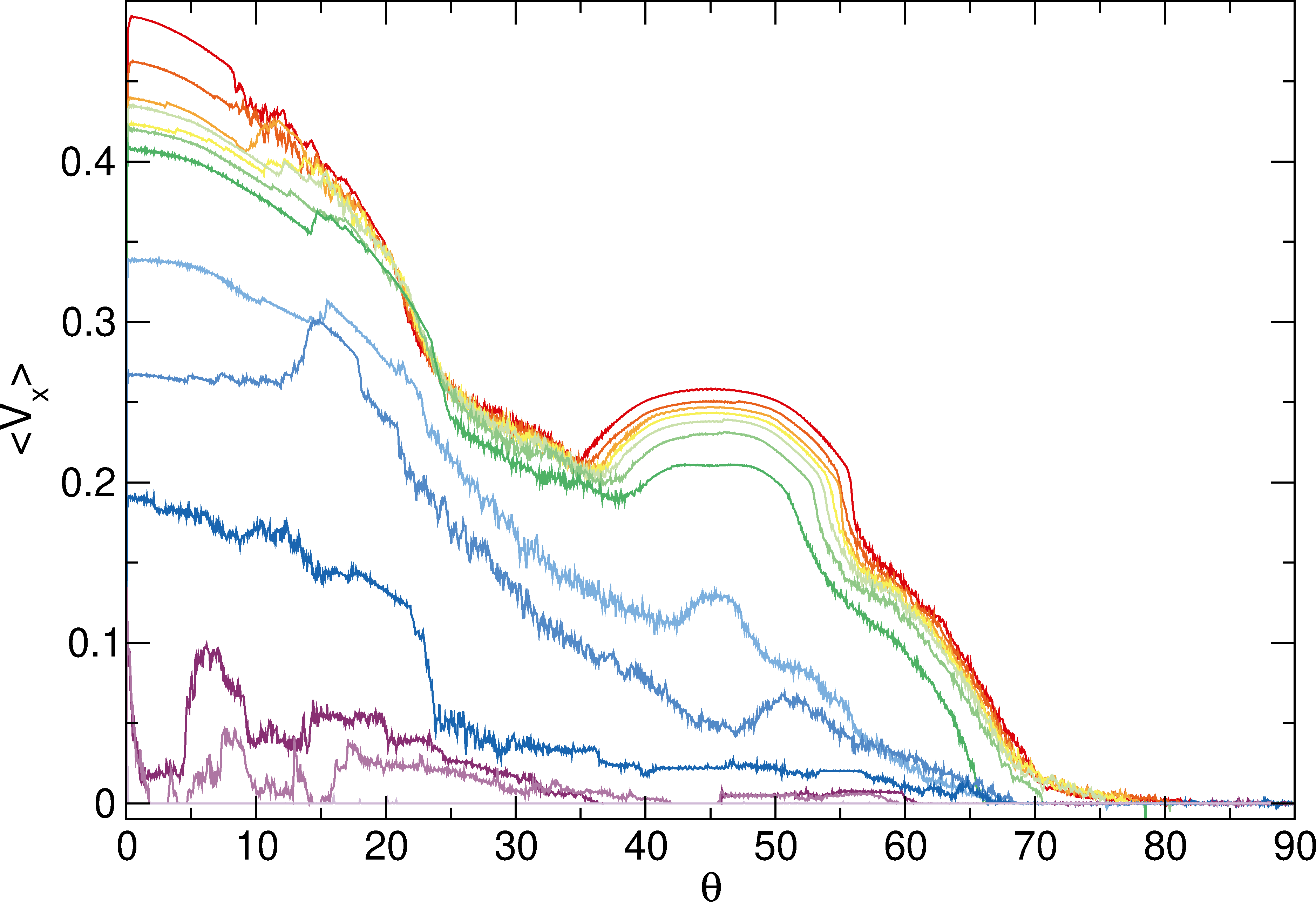}
\caption{
(a) $\langle V_{x}\rangle$ vs $\theta$ for the system
in Fig.~\ref{fig:13}
with $R_{d} = 0.5$, $a=4.0$, and
$\phi = 0.1934$
at $R_{\rm obs} =  1.45$, 1.475, 1.5, 1.5125, 1.525, 1.55, 1.575, 1.5875,
1.6,  1.6025, 1.6075, and $1.6375$, from top to bottom.
As $R_{\rm obs}$ increases, the system becomes clogged over a greater range
of driving angles.
At $R_{\rm obs} = 1.5875$ and $1.6$, directional locking still occurs,
while for $R_{\rm obs} = 1.6375$, 
there is complete clogging at every driving angle.
}
\label{fig:16}
\end{figure}

In Fig.~\ref{fig:16} we plot
$\langle V_{x}\rangle$ versus $\theta$ for the same system
in Fig.~\ref{fig:15}  at
$R_{\rm obs} =  1.45$, 1.475, 1.5, 1.5125, 1.525, 1.55, 1.575, 1.5875, 1.6,  1.6025, 
1.6075, and $1.6375$.
The velocity decreases with increasing $R_{\rm obs}$ and reaches zero 
for $R_{\rm obs} = 1.6375$.
The $p/q=1/1$ locking step is lost when $R_{\rm obs} \geq 1.575$. 
For $R_{\rm obs} = 1.6025$, the system reaches a 
clogged state near $\theta \approx 38^\circ$.
In this regime,
an increasing fraction of the sample contains
mobile disk configurations 
that block the flow.
When $R_{\rm obs} \geq 1.6125$,
the flow is blocked for all driving angles.

\begin{figure}
\includegraphics[width=\columnwidth]{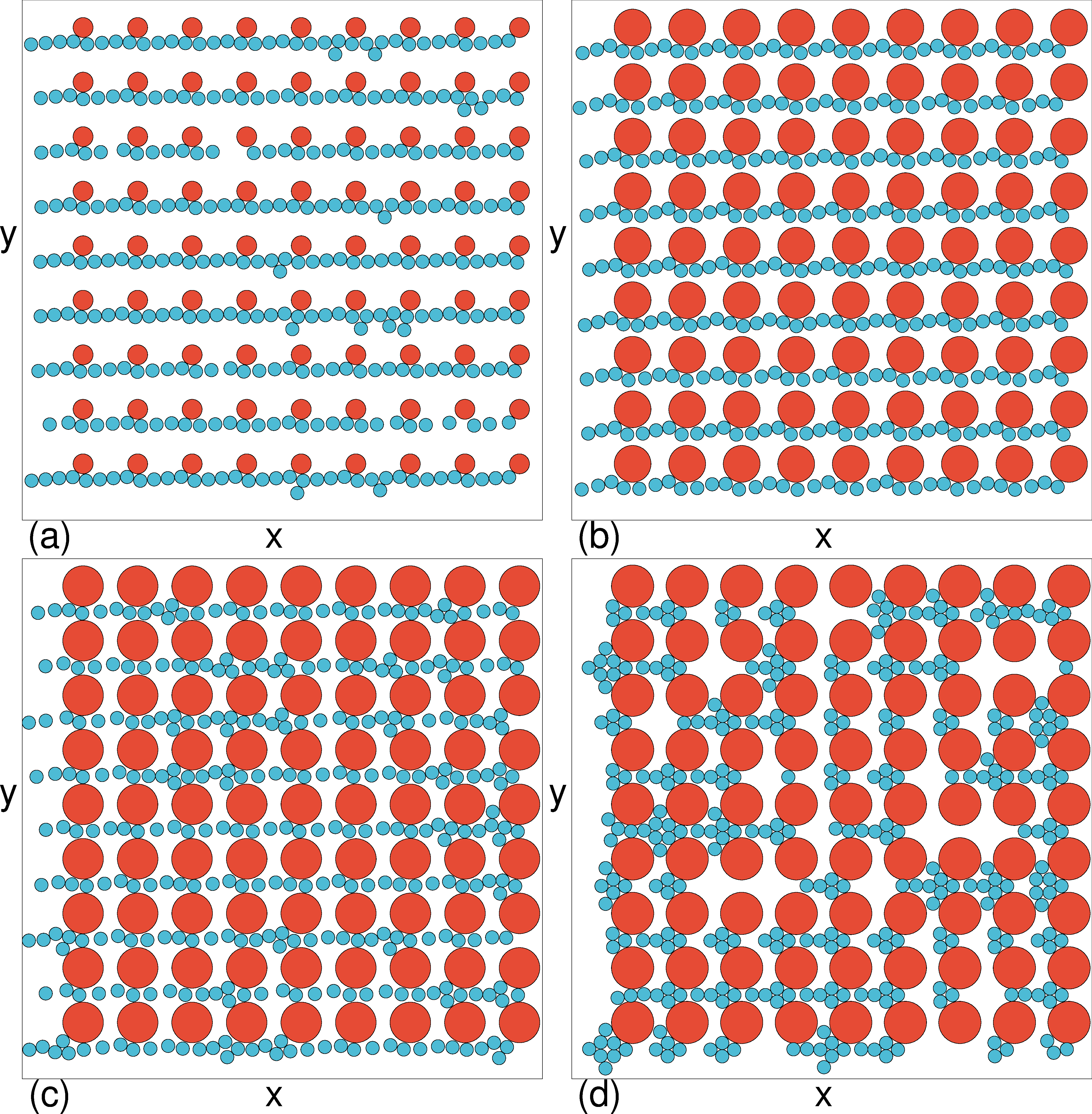}
\caption{
Obstacle (red) and disk (blue) locations for the system  
in Figs.~\ref{fig:15} and
\ref{fig:16}
with
$R_{d} = 0.5$, $a=4.0$,  and
$\phi = 0.1934$ in the $0^\circ$ locking phase.  
(a) $R_{\rm obs} = 0.75$.
(b) $R_{\rm obs} = 1.4$.
(c) $R_{\rm obs} = 1.55$, where partial clogging begins to occur.
(d) $R_{\rm obs} = 1.6125$, where there is a clogged state.
}
\label{fig:17}
\end{figure}

Figure~\ref{fig:17} illustrates the disk configurations
on the $0^\circ$ locking step for different obstacle sizes,
showing the evolution into a clogged state.
At $R_{\rm obs}=0.75$ in
Fig.~\ref{fig:17}(a), the disks form flowing 1D chains.
In Fig.~\ref{fig:17}(b)
at $R_{\rm obs} = 1.4$, the disks
interact more strongly with the
obstacles but no clogging occurs.
When $R_{\rm obs}=1.55$ as
in Fig.~\ref{fig:17}(c),
trimer configurations
form between adjacent obstacles,
intermittently blocking the flow. 
At $R_{\rm obs}=1.6125$ in Fig.~\ref{fig:17}(d),
the system is in a completely clogged state.
The local disk density is strongly heterogeneous in a clogged
sample, with some regions of high disk density accompanied by other
regions that contain no disks.
The clogged state we observe is similar to
that found for binary disks moving through
periodic obstacle arrays
\cite{Nguyen17}.    
For $R_{\rm obs}  = 1.6025$ and $1.6075$,
the clogging is directionally dependent and
the system does not clog
for flow along the
$x$, $y$, or $45^\circ$ directions,
but becomes completely blocked for flow at the other angles.

\begin{figure}
\includegraphics[width=\columnwidth]{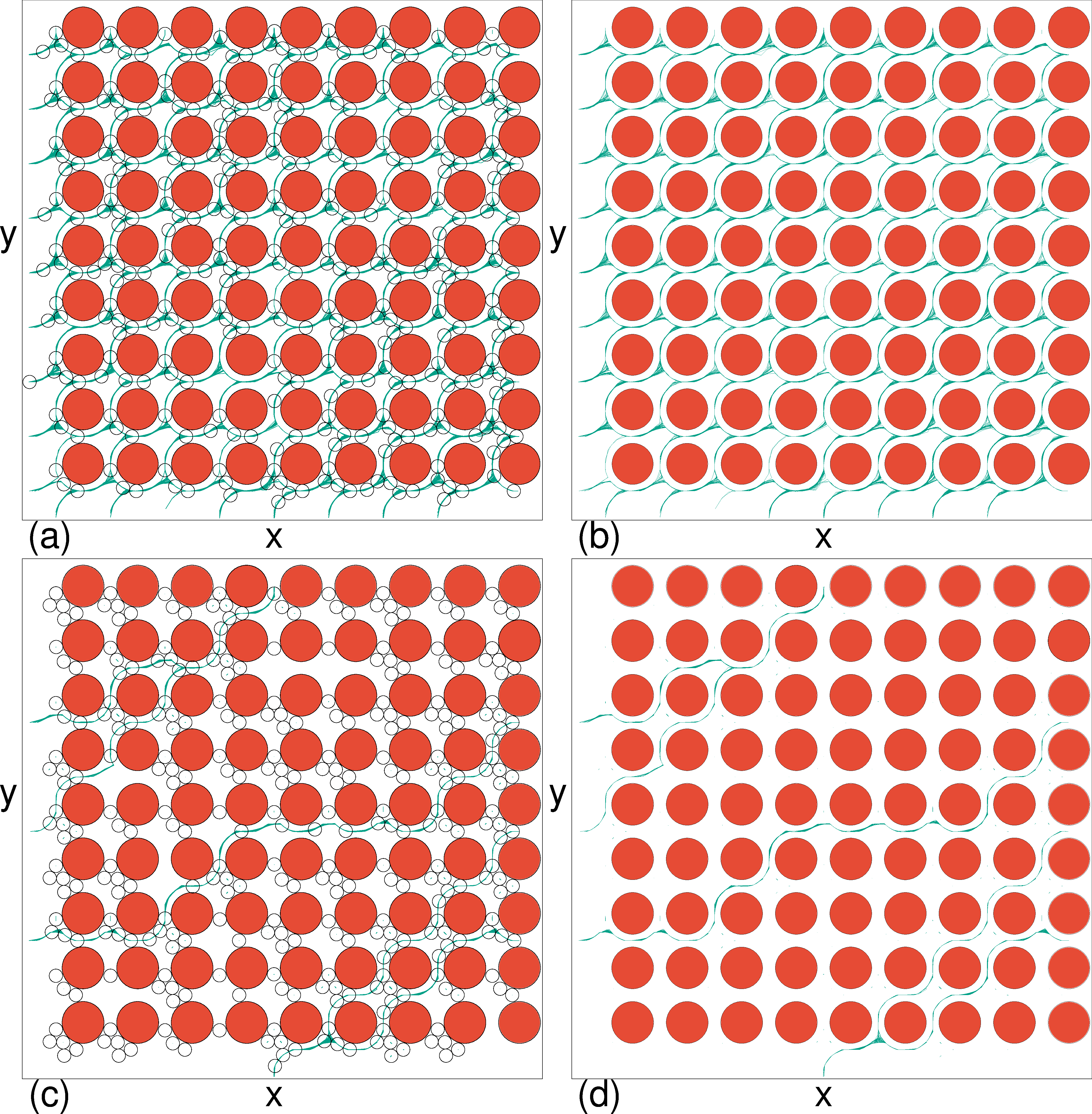}
\caption{
(a) Obstacle (red) and disk (open circle) locations along with the disk
trajectories (green)
for the system in Fig.~\ref{fig:16} with
$R_{d} = 0.5$, $a=4.0$, and
$\phi = 0.1934$
for $\theta = 40^\circ$.
(a) $R_{\rm obs}  = 3.15$. (b) The same as panel (a) showing only the obstacles
and the trajectories.
(c) $R_{\rm obs} = 3.2$. (d) The same as panel (c) showing only the obstacles
and the trajectories,
indicating that a partially clogged phase is present.       
}
\label{fig:18}
\end{figure}

In Fig.~\ref{fig:18}(a) we plot the obstacle and disk configurations
along with the trajectories
at $\theta = 40^\circ$ in
the system from Fig.~\ref{fig:16} with
$R_{\rm obs} = 1.575$, while in Fig.~\ref{fig:18}(b)
we show only the trajectories and obstacles.
The disks are beginning to accumulate
behind the obstacles but continue to flow around the obstacles,
as indicated in
Fig.~\ref{fig:18}(b).
The obstacles, disk configurations, and trajectories for the
same system with $R_{\rm obs}=1.6$ appear in 
Fig.~\ref{fig:18}(c), while
Fig.~\ref{fig:18}(d) shows only the
obstacles and trajectories.
The system is in a partially 
clogged phase containing
large regions where there is no flow
interspersed with some winding channels.
Those disks that continue to move channel predominantly along the
$0^\circ$ direction with occasional vertical jumping from one channel
to the next.

\begin{figure}
\includegraphics[width=\columnwidth]{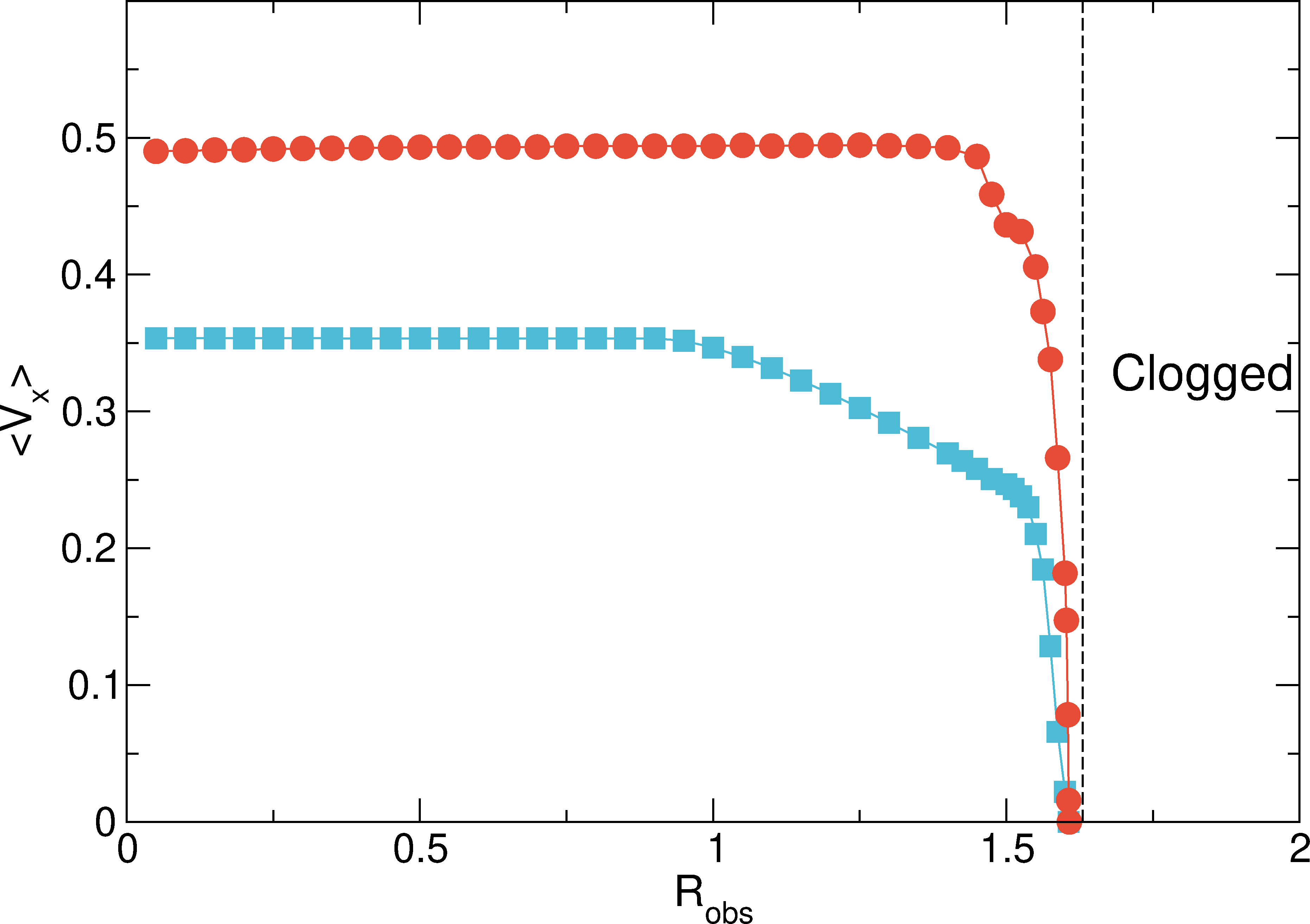}
\caption{
$\langle V_{x}\rangle$ vs $R_{\rm obs}$
for a system with $R_d=0.5$ and $a=4.0$
on the $0^\circ$ locking step at $\theta=3^\circ$ (red circles)
and on the $p/q=1/1$ locking step
at $\theta=45^\circ$ (blue squares), showing the crossover from 
a constant value to a clogged state.  
}
\label{fig:19}
\end{figure}

\begin{figure}
\includegraphics[width=\columnwidth]{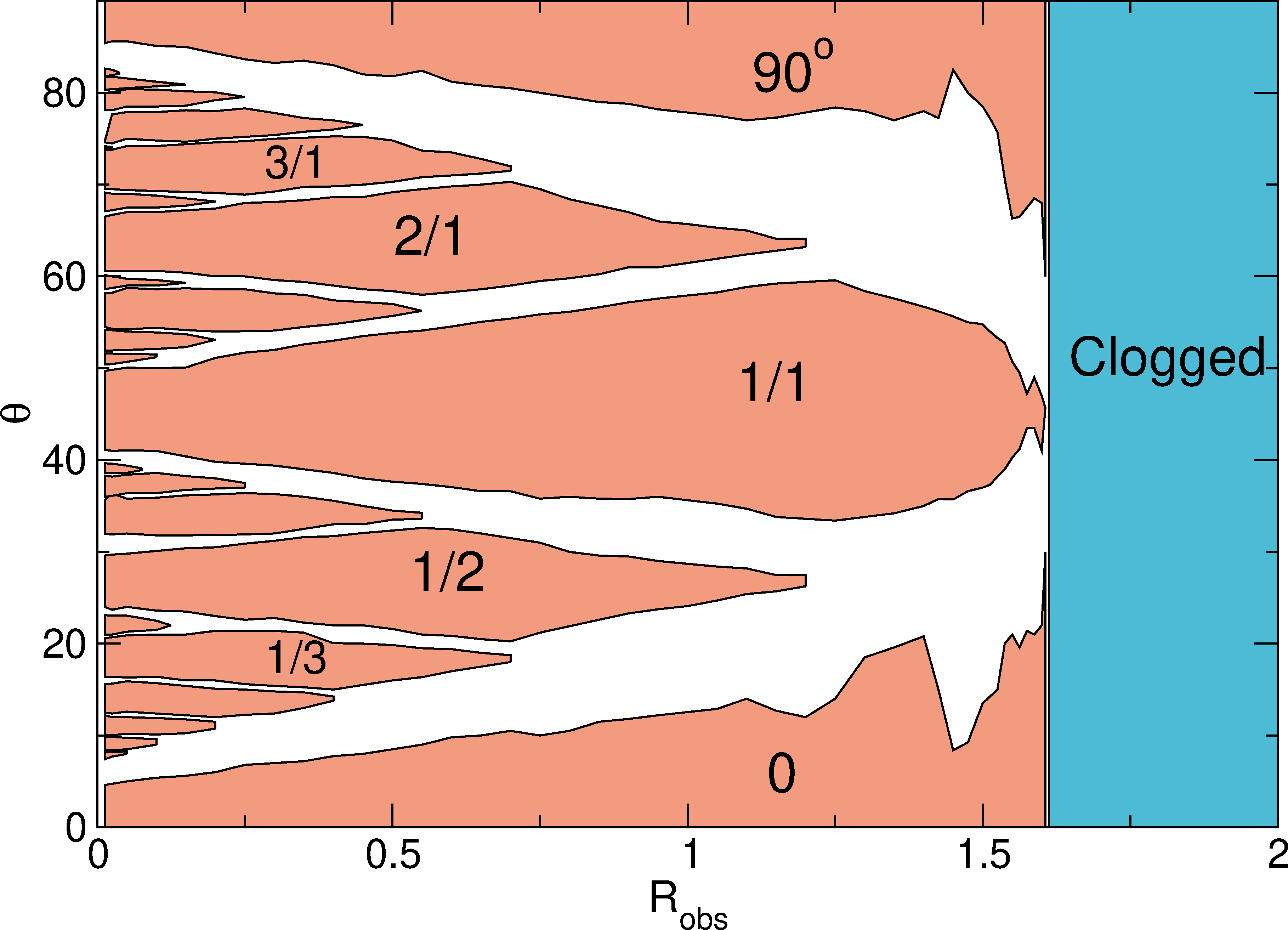}
\caption{
Locked regions (pink) and clogged region (blue) as a function of 
$\theta$
vs $R_{\rm obs}$ for the system in
Figs.~\ref{fig:15} and \ref{fig:16}
with
$R_d=0.5$ and $a=4.0$.
Locking steps with
$p/q = 0$, 1/3, 1/2, 1/1,
2/1, and $3/1$ are labeled along with the $90^\circ$ locking step.
}
\label{fig:20}
\end{figure}

The transition to a clogged state can
be quantified by measuring $\langle V_{x}\rangle$ as a function of
$R_{\rm obs}$ for a specific value of $\theta$.
In Fig.~\ref{fig:19} we plot $\langle V_{x}\rangle$ versus $R_{\rm obs}$
for $\theta = 3^\circ$ and $\theta= 45^\circ$.
When $\theta = 3.0^\circ$, the flow is locked along $0^\circ$ and the
velocity remains constant
up to $R_{\rm obs} = 1.45$, above which the
velocity drops until
reaching zero near $R_{\rm obs} = 1.61$. 
For $\theta = 45^\circ$, there is a smaller overall value of
$\langle V_x\rangle$ which remains
constant up to $R_{\rm obs} = 0.9$,
decreases linearly for
$0.9 < R_{\rm obs} < 1.54$, and then decreases more rapidly
before reaching zero 
near $R_{\rm obs} = 1.61$.
For other locking steps,
we find
a similar behavior
in which
$\langle V_{x}\rangle$ remains constant below a certain value of
$R_{\rm obs}$ 
before decreasing
linearly and then dropping rapidly to $\langle V_{x}\rangle = 0$
at higher $R_{\rm obs}$.
In Fig.~\ref{fig:20} we plot the step regions and clogged regions as
a function of
$\theta$ versus $R_{\rm obs}$
for the system in Fig.~\ref{fig:16}. 
We highlight the
$p/q = 0$, 1/3, 1/2, 1/1, 2/1, 3/1, and $90^\circ$ steps,
where the step width in $\theta$
generally grows with $R_{\rm obs}$
up to some critical value of $R_{\rm obs}$ before
decreasing again.
The higher order locking steps disappear
for $R_{\rm obs} > 0.5$.
The $0^\circ$ and $90^\circ$ locking steps diminish in width
near $R_{\rm obs} = 1.45$, which is correlated with
the onset of the partially clogged states.

We refer to the states in which the flow drops to zero as
clogged rather than jammed. Jamming
typically describes amorphous systems composed of 
loose particles such as grains, emulsions, or disks which
have no
quenched disorder
\cite{Liu98,Cates98,OHern03,Drocco05,Reichhardt14,Graves16}. 
In 2D systems,
jamming is typically associated with some type of long range growing
rigid correlation length
due to the build up of contact forces between the particles
\cite{Reichhardt14}. 
In a clogged system, the cessation of flow is more local and
is associated with individual bottlenecks \cite{Zuriguel15,Barre15}.
Clogging can occur for particles flowing through an individual
hopper when the particles adopt an arched configuration near the
mouth of the aperture.
The
susceptibility to clogging in this case
increases as the width of the aperture decreases.
In the clogging we observe,
the flow stops when the distance between
the obstacles is reduced due to an increase in
$R_{\rm obs}$, so the
system can be regarded as
a series of coupled hoppers.    

The clogging we find
is similar to the clogging phenomenon
studied in disordered systems with random 
obstacle arrays.
In the latter system,
a critical density of obstacles is required to block the flow
and the clogged states are inhomogeneous since
the individual blocked particles produce
higher density pileups behind them while
other regions of the sample contain few particles
\cite{Nguyen17,Reichhardt12a,Peter18,Stoop18}.
The clogged state forms for lower densities
when the quenched disorder is random compared to systems
containing periodic obstacle arrays
\cite{Reichhardt12a,Peter18,Stoop18}.
When the driving is applied along the $x$-direction with $\theta=0$,
the obstacles
in a periodic obstacle lattice with an easy flow channel aligned
in the $x$ direction must have a fairly large radius in order to induce
clogging,
while at larger driving angles,
the 
clogging is more similar to that found in
systems with random obstacles.
At finite temperature, the clogging is likely to 
be intermittent, with some clogged states
breaking up thermally and flowing for a period of time before reforming. 
The clogging can also be disrupted
by the application of an additional ac drive on top of the dc drive or 
by reversing the direction of the drive for a
period of time.
The clogging susceptibility also depends on the
magnitude of the driving force $F^{D}$ since
the disks are harmonic,
so that for weaker drives,
the system can reach a clogged state at lower $R_{\rm obs}$ and 
lower $\phi$. This property
will be studied in another work.    

\section{Varied Obstacle Lattice Constant}

\begin{figure}
\includegraphics[width=\columnwidth]{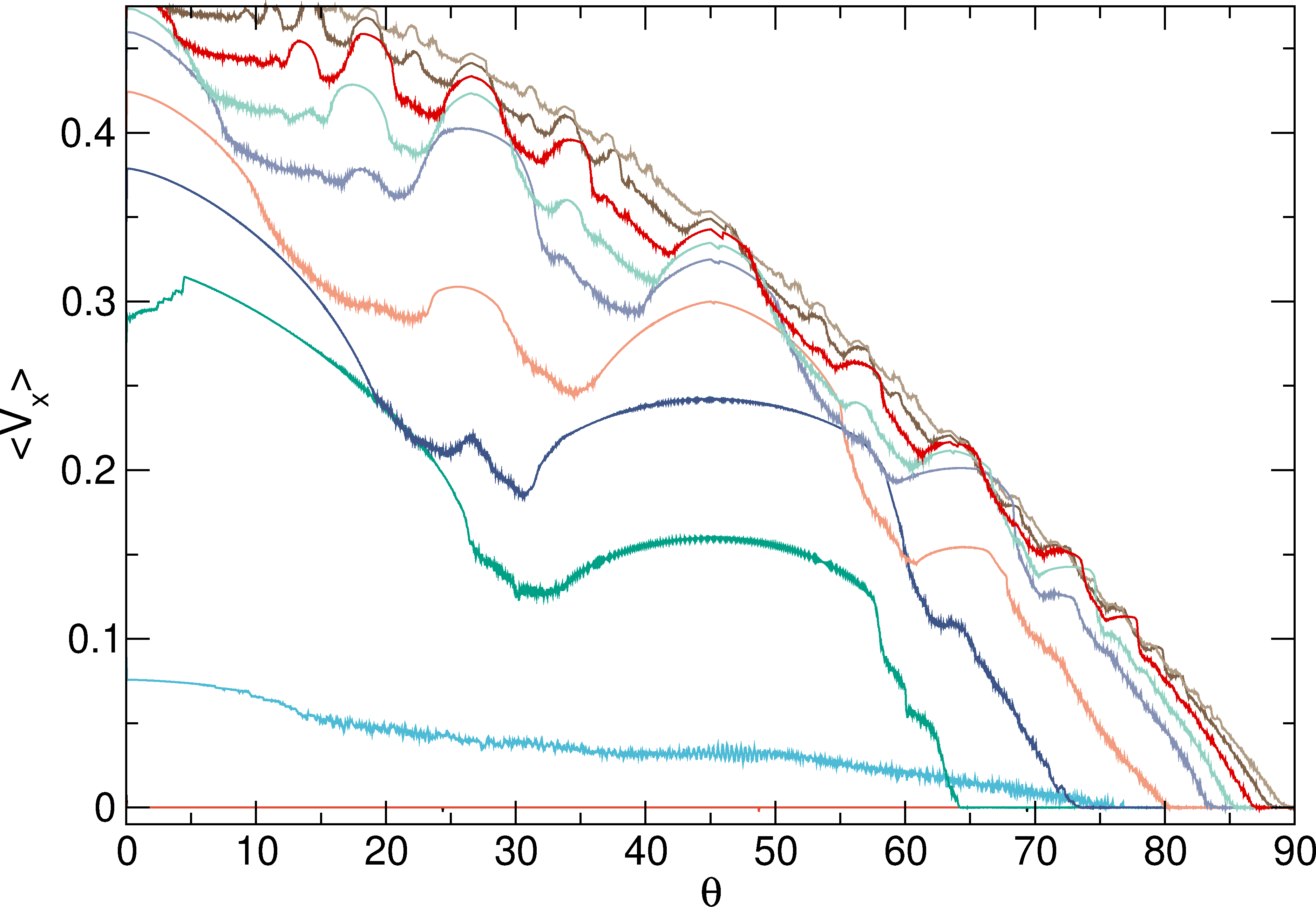}
\caption{
$\langle V_{x}\rangle$ vs $\theta$ for
samples with $N_d=400$, $R_{d} = 0.5$, and $R_{\rm obs} = 1.0$ at    
$a = 18$, 12, 9, 7.2, 6.0, 4.5, 3.6, 3.0, 2.5, and $2.25$,
from top to bottom.
At $a = 2.25$, the system is in a clogged state.
}
\label{fig:21}
\end{figure}

We next consider samples with fixed $R_{\rm obs}$ but varied $a$,
focusing on a system
with $N_{d} = 400$, $R_{d} = 0.5$, and $R_{\rm obs} = 1.0$.
In Fig.~\ref{fig:21} we plot $\langle V_{x}\rangle$ versus $\theta$ 
for $a = 18$, 12, 9, 7.2, 6.0, 4.5, 3.6, 3.0, 2.5, and $2.25$.
The velocity is highest
for the largest $a$
and the locking states
appear
as bumps.
As $a$ decreases,
some of the locking steps such as
those at $p/q=1/1$ and $p/q=1/2$ grow in width while
the higher order locking steps diminish in size.
The extent of the
$p/q=1/2$ locking step
begins to decrease when $a < 3.6$,
and for $a = 3.0$, only the $0^\circ$, $p/q=1/1$, and $90^\circ$ locking
phases appear.
The system
enters a partially clogged state
at $a = 2.5$,
and becomes
fully clogged for $a = 2.25$.
The evolution of the phases for decreasing $a$ is
similar to that found
for fixed $a$ and increasing
$R_{\rm obs}$,  
since in both cases the distance between the
surfaces of the obstacles decreases.

\begin{figure}
\includegraphics[width=\columnwidth]{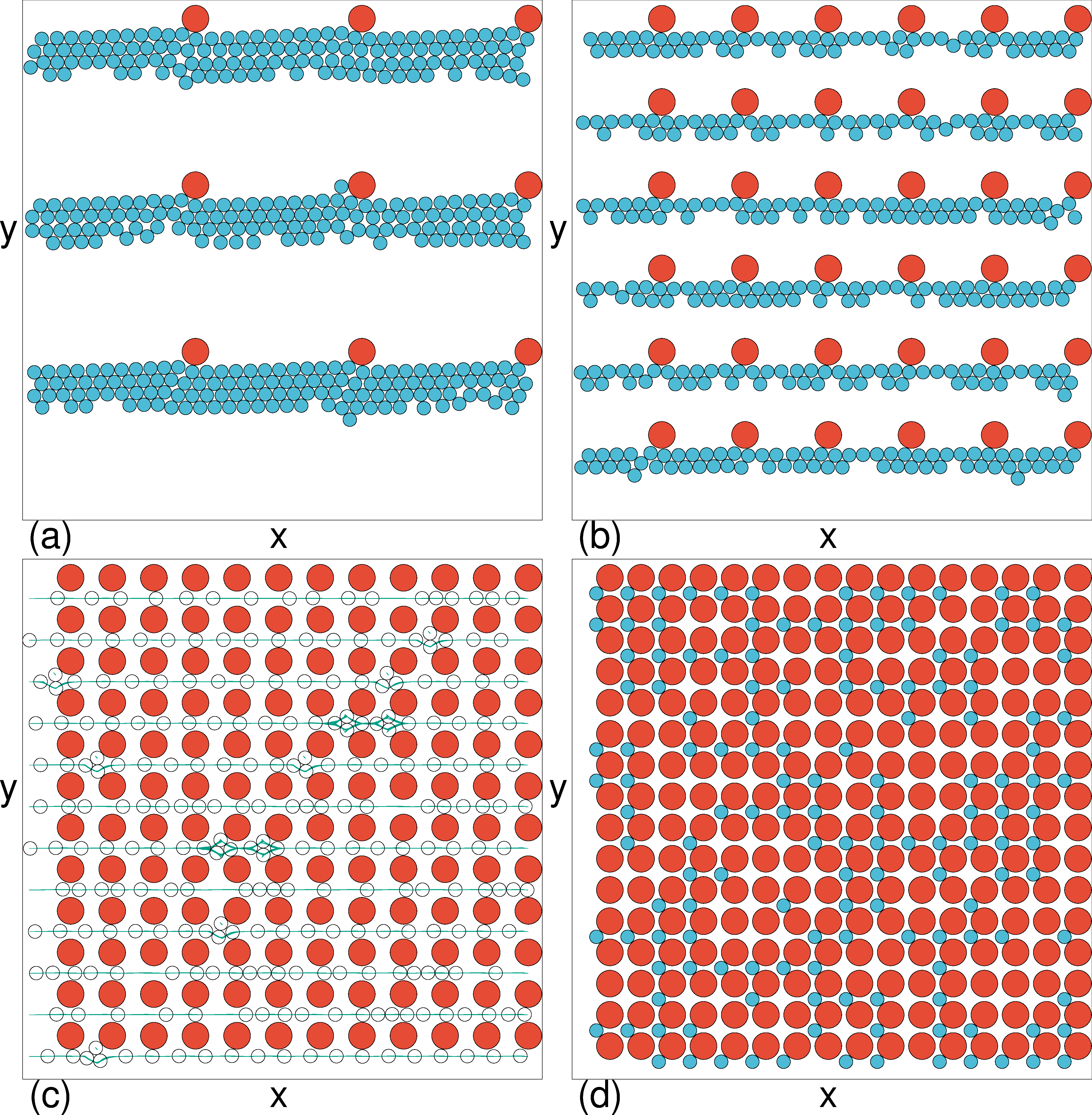}
\caption{Obstacle (red) and disk (blue or open circle) locations along
with the disk trajectories (green)
for the system in Fig.~\ref{fig:21}
with $N_d=400$, $R_d=0.5$, and $R_{\rm obs}=1.0$
on the $0^\circ$ locking step at a finite drive angle of $\theta = 2^\circ$.
(a) $a = 12$. (b) $a = 6.0$. (c) $a = 3.0$, showing
a partially clogged state.
(d) $a = 2.25$, where there is a fully clogged state.   
}
\label{fig:22}
\end{figure}

In Fig.~\ref{fig:22}(a) we illustrate
the disk configurations on the $0^\circ$ locking step
for a drive angle of $\theta = 2^\circ$ 
when $a = 12$,
where the system forms a density modulated stripe
containing between
between three and four rows of disks.
The stripes are pushed up against the obstacles due to the
finite angle of the drive.
Figure~\ref{fig:22}(b) shows the same
system at $a = 6.0$, where the
stripes are composed of between one and two rows of disks.
In Fig.~\ref{fig:22}(c) at $a = 3.0$,
the system forms a partially clogged state
with a single row of disks flowing between the obstacles
coexisting with
several regions in which a trimer disk arrangement
partially blocks the flow.
The fully clogged state at $a = 2.25$, shown in
Fig.~\ref{fig:22}(d),
occurs when the obstacles are so 
dense that individual disks cannot pass between them.
For higher $F^{D}$, the disks can
effectively depin and move between the obstacles,
which will be studied in another work.

\begin{figure}
\includegraphics[width=\columnwidth]{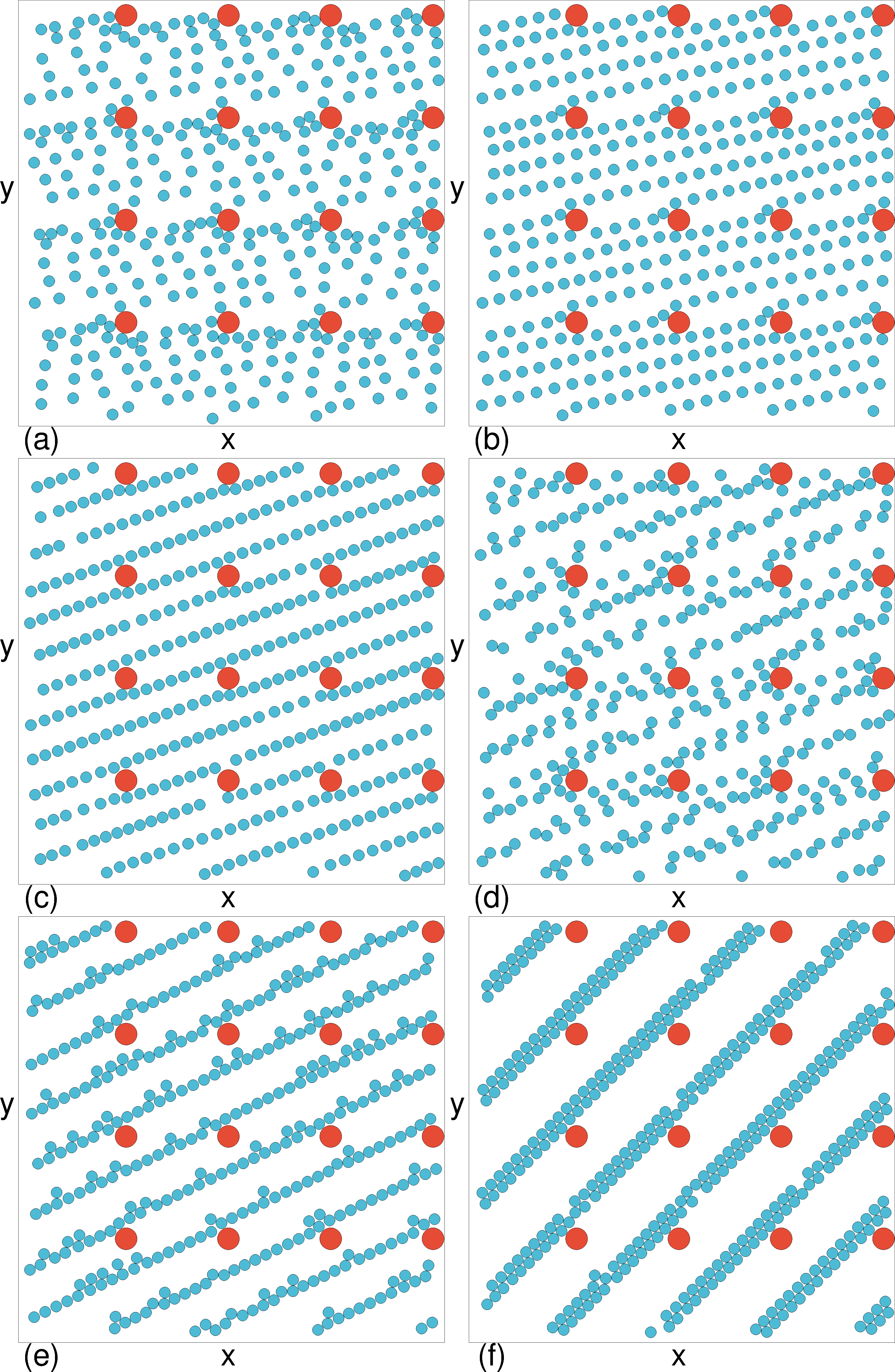}
\caption{Obstacle (red) and disk (blue) locations
for the system in Fig.~\ref{fig:21}
with $N_d=400$, $R_d=0.5$, and $R_{\rm obs}=1.0$ 
at $a = 9.0$. 
(a) $\theta = 10.5^\circ$, in a non-locking regime
where a periodic gradient in the disk density arises. 
(b) The $p/q=1/4$ locking phase.
(c) The $p/q=1/3$ locking phase.
(d) A non-locking phase between $p/q=1/3$ and $p/q=1/2$. 
(e) The $p/q=1/2$ locking step.
(f) The $p/q=1/1$ locking step where
a density modulated stripe structure appears.
}
\label{fig:23}
\end{figure}

\begin{figure}
\includegraphics[width=\columnwidth]{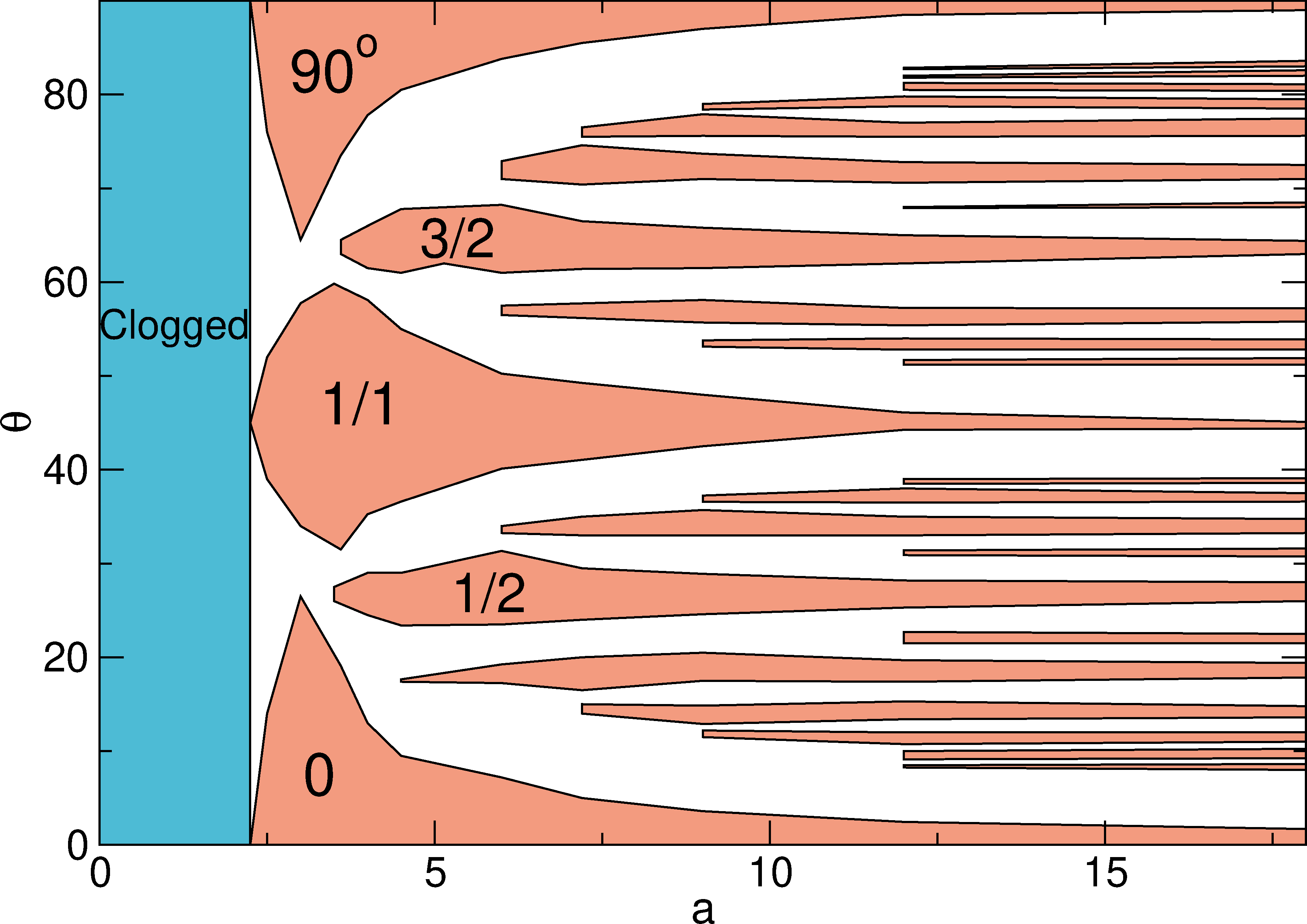}
\caption{
Locked regions (pink) and clogged region (blue) as
a function of $\theta$ vs $a$ for the system in
Fig.~\ref{fig:21} with $N_d=400$, $R_d=0.5$, and $R_{\rm obs}=1.0$.
Labels indicate the locations of
the $p/q=0$, 1/2, 1/1, 3/2, and $90^\circ$ locking steps.
Additional steps appear at 
$p/q = 1/7$, 1/6, 1/5, 2/5, 1/4, 1/3, 3/5, 4/5, 6/5, 4/3, 5/2, 3/1,
4/1, 5/1, 6/1, 7/1,
and $8/1$.     
}
\label{fig:24}
\end{figure}

At lower obstacle densities,
we find various types of pattern formation
at the locking phases, including
states with a density gradient.
In Fig.~\ref{fig:23}(a) we show the disk
and obstacle locations
for the system in Fig.~\ref{fig:21}
with $a = 9.0$ at $\theta = 10.5^\circ$ in a non-locking state.
The disks
form a partial square lattice in the regions between the obstacles,
while
a disordered pile up of disks forms immediately behind each obstacle.
On the $p/q=1/4$ locking step in Fig.~\ref{fig:23}(b) 
at $\theta = 18.7^\circ$,
the disks are more orderly and move
in a series of channels,
with a distortion in the rows produced by the deflection that occurs
when the disks collide with the obstacles.
In Fig.~\ref{fig:23}(c),
which shows the $p/q=1/3$ locking step,
the disks move in nearly straight lines
and undergo very few collisions with the obstacles.
Figure~\ref{fig:23}(d) shows the disordered configuration in
a non-locking regime
between the $p/q=1/3$ and $p/q=1/2$ locking steps.
On the $p/q=1/2$ step in
Fig.~\ref{fig:23}(e),
the particles form a density modulated phase,
while in Fig.~\ref{fig:24}(f)   
at the $p/q=1/1$ locking step,
there are two rows of disks moving at $45^\circ$ between the
obstacles. We observe several ordered
and disordered phases on the other locking steps.

In Fig.~\ref{fig:24} we indicate the locations of the locking steps
as a function of $\theta$ versus $a$ for the system in
Figs.~\ref{fig:21} to \ref{fig:23},
with labels denoting the
clogged
phase and the $p/q=0$, 1/2, 1/1, 3/2, and $90^\circ$ locking
steps.
Other locking steps
also appear for $p/q = 1/7$, 1/6, 1/5, 2/5, 1/4, 1/3, 3/5, 4/5, 6/5, 4/3, 5/2, 3/1, 4/1, 5/1, 6/1, 7/1,
and $8/1$.      
In systems that are larger in size than what we consider,
additional locking phases with smaller widths appear
at the larger values of $a$.  
For $2.25 < a \leq 3.0$,
partial clogging phases occur,
while for $a \leq 2.25$,
there is a complete clogging phase.
For $a >3.0$, the widths of the
$0^\circ$ and $90^\circ$ locking steps decrease
approximately as $1/a$.

\begin{figure}
\includegraphics[width=\columnwidth]{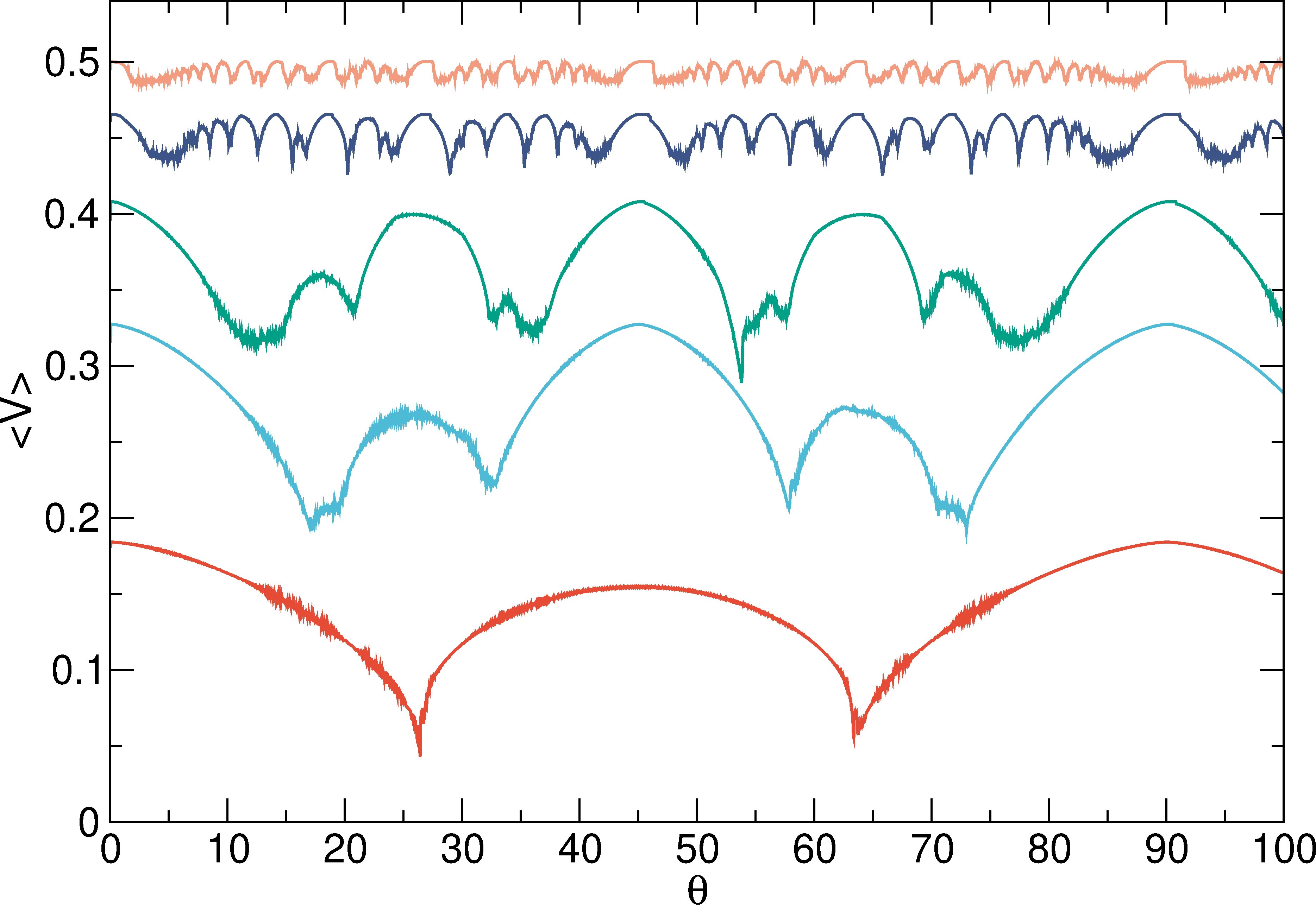}
\caption{
The net velocity $\langle V\rangle$ vs $\theta$
for
a system with $R_{\rm obs} = 0.5$, $R_d=0.5$, $N_d=400$, and
$a = 12$, 6, 4, 3, and $2.25$, from top to bottom.
}
\label{fig:25}
\end{figure}

\begin{figure}
\includegraphics[width=\columnwidth]{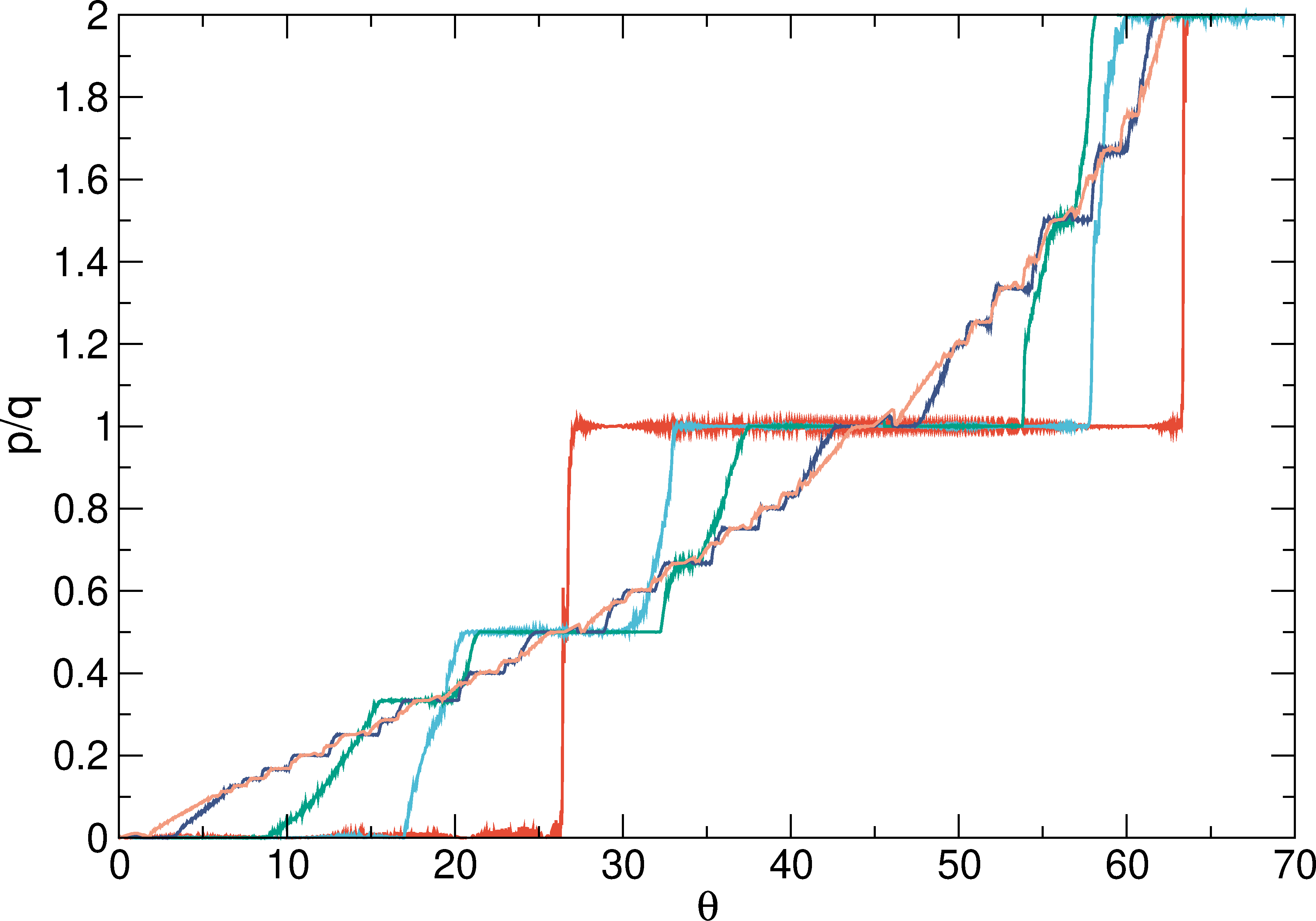}
\caption{
$p/q$ vs $\theta$ for the system in
Fig.~\ref{fig:25}
with $R_{\rm obs}=0.5$, $R_d=0.5$, and $N_d=400$
plotted over the range
$0 \leq \theta \leq 70^\circ$ and $0 \leq p/q \leq 2$
for $a=12$, 6, 4, 3, and 2.25, from upper left to lower left.
There are fewer, wider steps for smaller $a$.
}
\label{fig:26}
\end{figure}

In Fig.~\ref{fig:25} we plot the net velocity
$\langle V\rangle$ versus $\theta$ for a system with $N_d=400$,
$R_{\rm obs}=0.5$, and $R_d=0.5$ at
$a = 12$, 6, 4, 3, and $2.25$,
Here we have set $R_{\rm obs} = 0.5$ in order to access
higher disk densities.
For $a = 12$, only small dips appear in $\langle V\rangle$,
and for certain locking steps,
$\langle V\rangle=F^D=0.5$, indicating that there are
no collisions between the disks and the obstacles.
As $a$ decreases,
the overall velocity drops.
When $a = 4.0$,
only the $p/q=0,$ 1/3, 1/2, 1/1, 2/1, 3/1, and $90^\circ$
locking steps are present, while at $a = 3.0$, 
we find only the $p/q=0$, 1/2, 1/1, 2/1, and $90^\circ$ locking steps.
For $a = 2.25$, only the three most robust steps
of $p/q=0^\circ$, 1/1, and $90^\circ$ still appear.
For smaller $a$, the system reaches a completely clogged state.
In Fig.~\ref{fig:26} we plot $p/q$ versus $\theta$
for the system in Fig.~\ref{fig:25}
showing the growth of the locking phase step widths with 
decreasing $a$.

\begin{figure}
\includegraphics[width=\columnwidth]{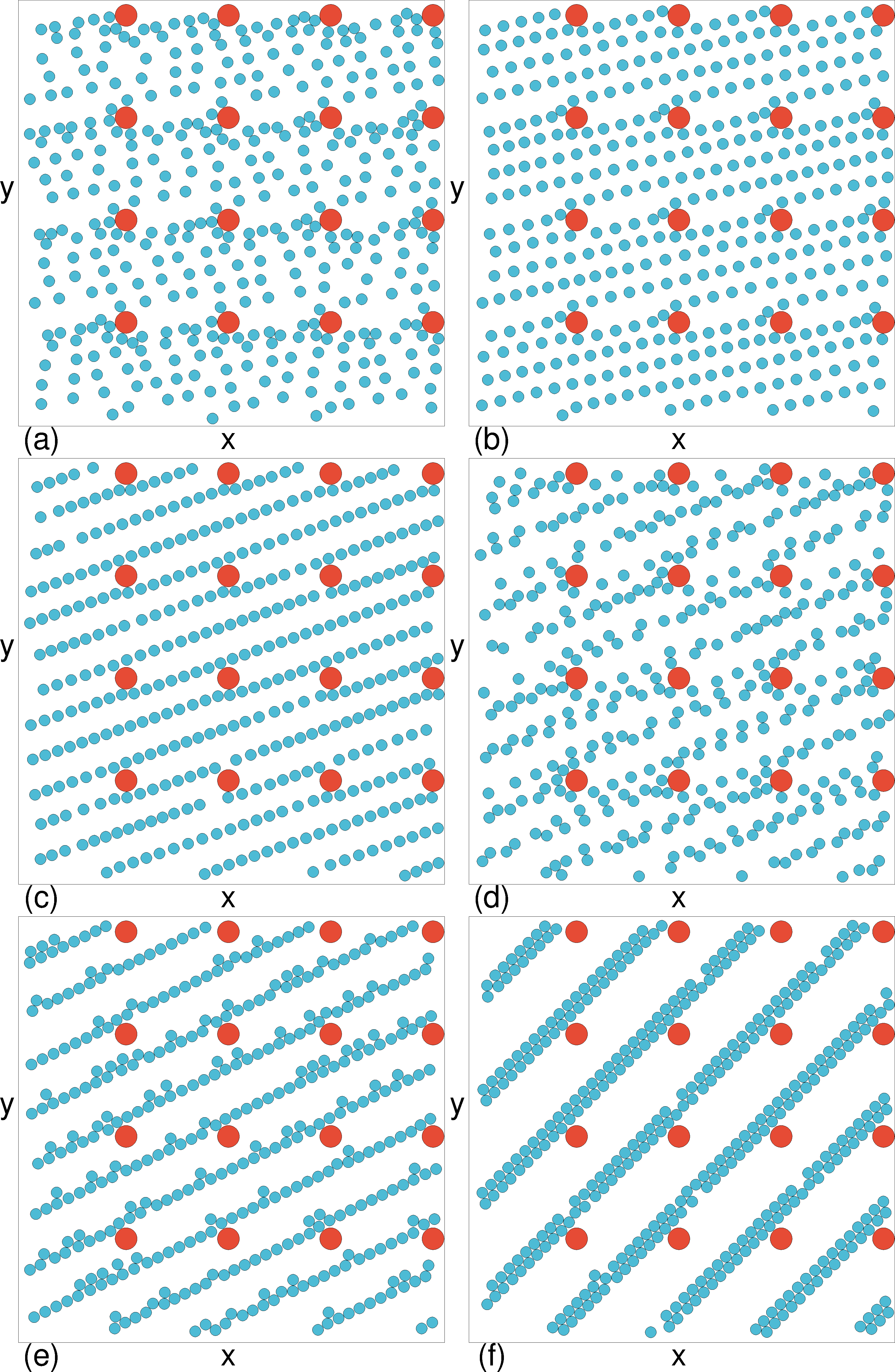}
\caption{Obstacle (red) and disk (blue) positions for 
  a system with $a = 6.0$, $R_d=0.5$,
$R_{\rm obs}=1.0$,
and density $\phi = 0.525$ showing pattern formation.
(a) The $0^\circ$ locking state.
(b) $p/q = 1/5$ locking.
(c) $p/q = 1/3$ locking.
(d) A non-locking phase just below the $p/q=1/2$ step.
(e) $p/q = 1/1$ where the system is ordered with two rows of disks.
(f) $p/q = 6/1$ showing a density
modulated disk arrangement.
}
\label{fig:27}
\end{figure}

We have also examined samples with
larger $a$ and higher disk densities,
and find density modulated states similar to those
described above.
In Fig.~\ref{fig:27} we show the disk configurations at
$a = 6.0$, $R_d=1.0$,
$R_{\rm obs} = 1.0$, 
and disk density $\phi = 0.525$.
Figure~\ref{fig:27}(a) illustrates the $0^\circ$ locking step where
a disordered stripe phase appears.
For larger $\theta$ that is still within the
$0^\circ$ locking regime,
the stripes become 
more compact and contain only four rows of disks.  
In Fig.~\ref{fig:27}(b) we plot the disk configuration on
the $p/q = 1/5$ locking step, while in Fig.~\ref{fig:27}(c) we show the
configuration at the $p/q = 1/3$ locking.
Figure~\ref{fig:27}(d) gives an example of the configuration
in the non-locking regime just below the $p/q=1/2$ locking step, while
Fig.~\ref{fig:27}(e) shows the locking at $p/q = 1/1$
where the system forms an ordered state containing
two rows moving at $45^\circ$.
In Fig.~\ref{fig:27}(f) we illustrate the density modulated state
which forms at
$p/q = 6/1$.

\section{Summary}

We have examined the directional locking for a
collection of disks moving through a square 
obstacle array, where we vary the mobile disk density,
the obstacle radius, and the obstacle lattice constant.
We find
strong collective effects which produce
a rich variety of patterns.
On the steps with strong directional locking, such as $45^\circ$,
the disks can form linear chains containing one or more rows.
For other locking directions,
we find square moving lattices or density modulated states.
The disk trajectories on the locking steps form ordered patterns and
the disks move elastically without exchanging neighbors.
In the non-locking regimes,
disordered or liquid-like states appear in which
the disk trajectories mix.
On the directional locking steps,
the disk velocities are not fixed
but form a parabolic shape with minima 
at the transitions into and out of the locked phase.
In contrast, the ratio $p/q$ describing the direction of motion is  
constant on each locking step.
As the disk density increases, the number of possible locking phases
diminishes
due to the increasing frequency of disk-obstacle collisions
which makes it impossible for the disks to form a collectively moving
pattern at certain drive angles.
When the obstacle radius becomes larger,
the number of locking steps decreases
and the system first reaches a partially clogged phase in which a portion
of the disks are stationary before entering
a fully 
clogged state.
For large obstacle lattice constants,
a variety of moving stripe or density 
modulated states appear on the locking steps,
and the number of locking phases decreases with decreasing
obstacle lattice
constant until the system reaches a completely clogged state.
Our results should be relevant to bubbles, 
emulsions, uncharged colloids, and magnetic textures moving through
obstacle arrays, and they suggest a new way to 
dynamically generate stripe and density modulated phases.      

\acknowledgments
This work was supported by the US Department of Energy through
the Los Alamos National Laboratory.  Los Alamos National Laboratory is
operated by Triad National Security, LLC, for the National Nuclear Security
Administration of the U. S. Department of Energy (Contract No.~892333218NCA000001).

\bibliography{mybib}

\end{document}